\pdfoutput=1
\documentclass[useAMS,usenatbib]{mn2e}
\usepackage{graphicx}
\usepackage{amssymb}
\usepackage{lscape}
\usepackage{rotating}
\usepackage{subfigure}
\usepackage{amsmath}

\def\deg{\ifmmode^\circ\else$^\circ$\fi}

\def\Q{\ifmmode\mathcal{Q}\else$\mathcal{Q}$\fi}
\def\Mach{\ifmmode\mathcal{M}\else$\mathcal{M}$\fi}

\title[Star formation around bubble CN 148]
{Star formation around the mid-infrared bubble CN 148}

\author[L.~K. Dewangan, D.~K. Ojha, J.~M.~C. Grave, \& K.~K. Mallick]
{L.~K. Dewangan$^{1}$\thanks{Lokesh.Dewangan@astro.up.pt}, D.~K. Ojha$^{2}$\thanks{ojha@tifr.res.in}, J.~M.~C. Grave$^{1,3}$\thanks{jgrave@astro.up.pt}, \& K.~K. Mallick$^{2}$\thanks{kshitiz@tifr.res.in}\\
$^{1}$Centro de Astrof\'{\i}sica, Universidade do Porto, Rua das Estrelas, 4150-762 s/n Porto, Portugal.\\
$^2$Department of Astronomy and Astrophysics, Tata Institute of Fundamental Research, Homi Bhabha Road, 
Mumbai 400 005, India.\\
$^3$FCNET, Universidade Lus\'{o}fona do Porto, Rua Augusto Rosa 24, 4000-098, Porto, Portugal}

\begin{document}

\date{ }

\pagerange{\pageref{firstpage}--\pageref{lastpage}} \pubyear{2014}

\maketitle

\label{firstpage}

\begin{abstract} 
We present a multi-wavelength study to analyse the star formation process associated with the 
mid-infrared bubble CN 148 (H\,{\sc ii} region G10.3-0.1), which harbors an 
O5V-O6V star. The arc-shaped distribution of molecular CO(2-1) emission, 
the cold dust emission, and the polycyclic aromatic hydrocarbon features trace a 
photodissociation region (PDR) around the H\,{\sc ii} region. We have identified 371 
young stellar objects (YSOs) in the selected region and, interestingly, their spatial 
distribution correlates well with the PDR. 41\% of these YSOs are present in 13 clusters, 
each having visual extinction larger than 16 mag. The clusters at the 
edges of the bubble (both northeast and southwest) are found to be relatively younger than 
the clusters located further away from the bubble. 
We also find that four 6.7 GHz methanol masers, two Extended Green Objects, an ultra-compact H\,{\sc ii} region, and a 
massive protostar candidate (as previously reported) are spatially positioned at the edges of 
the bubble. The existence of an apparent age gradient in YSO clusters and different 
early evolutionary stages of massive star formation around the bubble suggest 
their origin to be influenced by an H\,{\sc ii} region expanding into the surrounding 
interstellar medium. The data sets are suggestive of triggered star formation.
\end{abstract}
\section{Introduction}
\label{sec:intro} 
\begin{figure*}
\includegraphics[width=16cm]{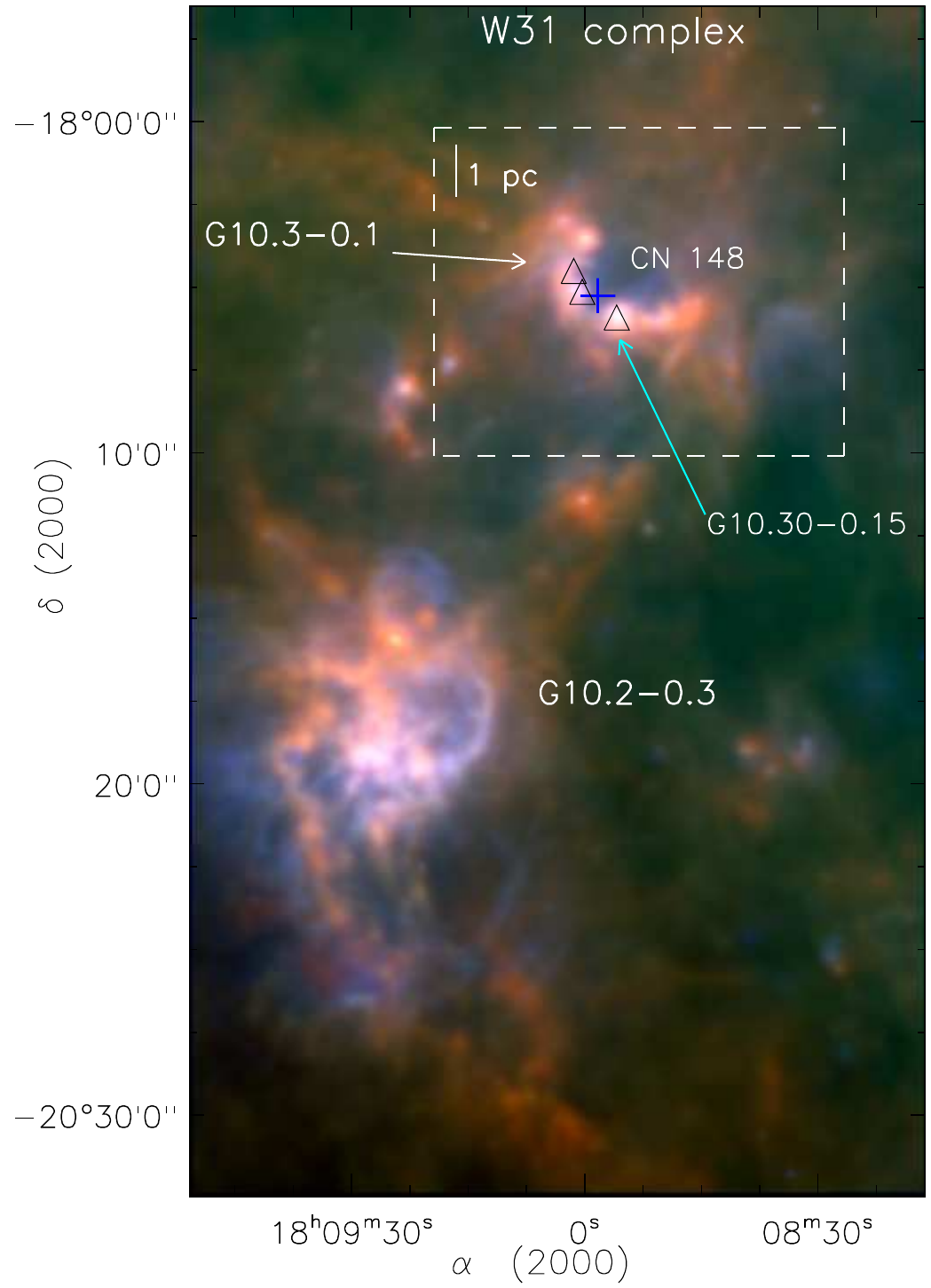}
\caption{{\it Herschel} Hi-GAL continuum images (350 $\mu$m (red), 160 $\mu$m (green), and 70 $\mu$m (blue)) in log scale of the W31 region.
The sub-regions G10.2-0.3 and G10.3-0.1 of the W31 complex are labeled in the figure. 
The region selected in this paper is highlighted by the white dashed box in figure.
The positions of IRAS 18060-2005 (+) and three 6 cm radio detections ($\triangle$)
are also marked in the selected region. The scale bar on the top left in the G10.3-0.1 region shows a size of 1 pc at a distance of 2.2 kpc.}
\label{fig1u}
\end{figure*}

The mid-infrared (MIR) bubble, CN 148 \citep{churchwell07} is part of an extended radio 
H\,{\sc ii} region, G10.3-0.1 \citep[spatial extent $\sim$ 12$\farcm$8 $\times$ 4$\farcm$6;][]{kim01}, and 
contains the IRAS 18060-2005 source. 
The H\,{\sc ii} region is thought to be associated with the 
larger W31 molecular cloud complex \citep[e.g.,][]{reich90}. The W31 complex is comprised of two extended H\,{\sc ii} regions, 
G10.2-0.3 and G10.3-0.1 (see Figure~\ref{fig1u}). 
The G10.3-0.1 region is estimated to have 
a far-infrared (FIR) luminosity of about 10$^{5}$\,L$_{\odot}$ \citep{ghosh89,wood89}. 
The object CN 148 was classified as a complete 
or a closed-ring with an average radius (diameter) and thickness of 1$\farcm$41 (2$\farcm$82) 
and 0$\farcm$65 \citep{churchwell07}, respectively. 
The MIR bubble gains its importance due to two embedded sources, 
``18060nr1733" and ``18060nr2481", which were classified as O5V-O6V and 
O8V-B2.5V spectral type objects, respectively, based on near-infrared (NIR) {\it K}-band 
spectra \citep{bik05}. 
\citet{bik05} also found direct spectroscopic signatures of strong winds, associated 
with O5-O6 stars, to be present in the ``18060nr1733" {\it K}-band spectra. 
The H\,{\sc ii} region G10.3-0.1 has been included in numerous studies of the larger W31 complex, carried 
out at multiple wavelengths \citep{becker94,brand94,walsh98,kim01,kim02,beuther11}. 
The 21 cm radio continuum emission \citep[spatial resolution $\sim$ $37\arcsec \times 25\arcsec$; from][]{kim01} 
revealed an extended bipolar morphology for the G10.3-0.1 region, using the Very Large Array (VLA) DnC-array. 
\citet{kim01} also reported an ultra-compact (UC) H\,{\sc ii} region G10.30-0.15 and 
another compact component surrounded by extended bipolar ionized envelopes. 
The 21 cm flux was also used to classify the ionizing source 
as an O4.5 star, assuming a distance of 6.0 kpc. 
\citet{kim02} suggested the champagne flow model to 
explain the bipolar radio morphology, using the molecular CO and CS 
line observations (at $60\arcsec$ resolution). These authors found a very 
high $^{12}$CO J = 2-1/J = 1-0 intensity ratio (average value $\sim$ 1.4) 
around the bubble (referred to as a ``hollow" in their paper), which led them to suggest ongoing 
massive star formation (MSF) around the H\,{\sc ii} region. 
Using the champagne model, \citet{kim02} estimated the dynamical age of $\sim$ 1.2 Myr for the H\,{\sc ii} region G10.3-0.1.  
They concluded that the H\,{\sc ii} region and its ionizing star is interacting with the surrounding ambient 
material. Subsequent radio observations at a higher spatial resolution, 
mapping of the $^{13}$CO(2-1) and C$^{18}$O(2-1) emission (resolution $\sim$ 27$\farcs$5), 
and the detection of 870 $\mu$m APEX ATLASGAL \citep{schuller09} dust continuum emission in this region, revealed 
several high-mass clumps \citep{beuther11}.  
\citet{beuther11} reported three different evolutionary stages of star formation, mainly around the central part of the bipolar nebula
in G10.30-0.1. These authors also argued that a triggered star formation process may be responsible 
for the different evolutionary stages in this region. 

Apart from the above characteristics, G10.30-0.1 is known to be 
associated with several masers and Extended Green Objects (EGOs). Three 6 cm (5 GHz) VLA 
radio detections \citep{becker94}, one water maser \citep{brand94}, and four 6.7 GHz Class~II methanol 
masers \citep{walsh98} were reported in this region. Two GLIMPSE EGOs 
\citep{cyganowski08} were also observed in the northeast (G10.34-0.14) 
and southwest (G10.29-0.13) directions and both these EGOs are associated 
with 6.7 GHz Class~II as well as 44 GHz Class~I methanol masers. 
The EGOs are thought to be massive YSOs (MYSOs), still undergoing the accretion phase \citep{cyganowski09}. 
Different values of the distance (2.2 kpc, 3.4 kpc, 6.0 kpc, and 14.5 kpc) to G10.3-0.1 are
found in the literature \citep[e.g.][]{wilson74,downes80,corbel97,kim01,kim02,bik05,cyganowski09,moises11,beuther11,wienen12}. 
\citet{bik05} estimated the kinematic distances of the two sources (``18060nr1733" and ``18060nr2481") in the range 2.1 (near) - 14.6 (far) kpc, 
using the CS molecular line radial velocities combined with a Galactic rotation model, while \citet{moises11} estimated an average 
{\it K}-band spectrophotometric distance of $\sim$ 2.39 kpc. 
The velocities of the molecular gas \citep[10--15 km s$^{-1}$;][]{kim02}, the ionized gas \citep[8--12 km s$^{-1}$;][]{downes80,kim01}, 
four 6.7 GHz methanol masers \citep[13--20 km s$^{-1}$;][]{walsh98}, and 
two EGOs \citep[12.7--13.6 km s$^{-1}$;][]{cyganowski09} suggest that they are part of the same star-forming region. 
\citet{wienen12} reported observations of three ammonia (NH$_{3}$) inversion transitions 
and estimated the radial velocities of the NH$_{3}$ gas between 11.3 and 14.5 km s$^{-1}$. 
\citet{cyganowski09} and \citet{wienen12} reported the near kinematic distance as 1.9 kpc and 2.2 kpc, respectively. 
Considering the fact that the ionizing candidates, methanol masers, and EGOs are physically associated with the G10.3-0.1 
region, the distance of 2.2 kpc is consistent. 
The spectral type estimated from the 21 cm flux at a distance of 2.2 kpc is also in agreement with the 
spectral type determined using the {\it K}-band spectra for the sources (see Section~\ref{subsec:lymnflx} in this paper). 
Therefore, we have chosen a distance of 2.2 kpc for the bubble CN 148 and have adopted this 
value throughout the present work. Considering the near kinematic distance range \citep[1.9--2.5 kpc;][]{bik05,cyganowski09,wienen12}, 
we anticipate the resulting uncertainty in the chosen distance to be $\sim$ 15\% (i.e., 0.3 kpc).

All these studies clearly indicate the presence of ongoing star formation and of massive stars in a very early evolutionary stage. 
Previous studies also demonstrate the influence of the H\,{\sc ii} region on the surrounding interstellar medium (ISM). 
These multiple signposts of star formation as well as the interesting characteristics of the region have motivated a further 
investigation of CN 148. To this end, we revisited the source using {\it Spitzer} Infrared Array Camera \citep[IRAC;][]{fazio04} data in 
conjunction with unpublished high sensitivity UKIRT NIR data, a narrow-band H$_{2}$ v= 1-0 S(1) 2.12 $\mu$m image, 
and a radio continuum map at 20 cm of the wider area around the bubble. 
While systematically taking care of all possible contaminants, we have studied 
the embedded young stellar populations, their relative distribution with 
respect to the molecular and ionized gas, cold dust, and polycyclic 
aromatic hydrocarbon (PAH) emission. 

A variety of infrared (IR) and radio (archival and literature) data used for this purpose are 
described, along with the analysis methods, in Section~\ref{sec:obser}. 
In Section~\ref{sec:data}, we present the results of the combined analysis. 
The possible star formation scenario is discussed in 
Section~\ref{sec:data1}. The main conclusions are summarized in Section~\ref{sec:conc}.
\section{Data and analysis}
\label{sec:obser}
To study the embedded populations as well as the ongoing star formation 
process in the target region, we used NIR and MIR surveys. 
The size of the selected field is $\sim 12$\farcm$4  \times 9$\farcm$9$, 
centered at $\alpha_{2000}$ = 18$^{h}$08$^{m}$53$^{s}$,
$\delta_{2000}$ = -22$\degr$05$\arcmin$08$\arcsec$.

NIR {\it JHK} images and the point source catalog 
were obtained from the UKIDSS 6$^{th}$ archival data release
(UKIDSSDR6plus) of the Galactic Plane Survey (GPS) \citep{lawrence07}. 
UKIDSS observations were made using the UKIRT Wide Field Camera
\citep[WFCAM;][]{casali07} and fluxes were calibrated using the Two Micron
All Sky Survey data \citep[2MASS;][]{skrutskie06}.  
WFCAM is an array of four $2048 \times 2048$ pixels Rockwell Hawaii-II arrays with a pixel size of 0$\farcs$4. 
The details of basic data reduction and calibration procedures are described in
\citet{dye06} and \citet{hodgkin09}, respectively. Only reliable NIR
sources were extracted from the publicly access catalog, following
the recommendations from \citet{lucas08}. Sources visible in all the
three NIR ({\it JHK}) bands and those visible only in the {\it H} and {\it K}
bands were separately extracted for our selected region. 
The adopted criteria allowed for the removal of saturated sources, non-stellar
sources, and unreliable sources near the sensitivity limits. Magnitudes
of saturated bright sources were obtained from 2MASS and appended to
the resulting catalogs. Only those sources with magnitude error of 0.1
or less in each band were selected for the study to ensure good
photometric quality.  As a result, 7864 sources were obtained, which
were common to all the three (J, H, and K) bands. Additionally 4710
sources were found in the  H and K bands, without a {\it J} counterpart.
 
IR data from the {\it Spitzer} IRAC ch1 (3.6 $\mu$m), ch2 (4.5 $\mu$m), ch3 (5.8  $\mu$m), and ch4 (8.0 $\mu$m) 
bands were obtained from the GLIMPSE survey \citep{benjamin03,churchwell09}, while the Multiband Imaging 
Photometer for {\it Spitzer} (MIPS) image at 24 $\mu$m was obtained from the MIPS Inner Galactic Plane 
Survey \citep[MIPSGAL;][]{carey05}. 
Photometry was extracted from the GLIMPSE-I Spring '07 highly reliable Point-Source
Catalog. Some sources are well detected in the images but do not have
photometric magnitudes in the GLIMPSE-I catalog. We obtained aperture
photometry for such sources using the GLIMPSE images at a plate scale
of 0$\farcs$6/pixel. The photometry was done using a
2$\farcs$4 aperture radius and a sky annulus from 2$\farcs$4 to
7$\farcs$3 in IRAF\footnote[1]{IRAF is distributed by the 
National Optical Astronomy Observatory, USA}. 
Apparent magnitudes were calibrated using the 
IRAC zero-magnitude flux densities, including aperture 
corrections listed in the IRAC Instrument Handbook-Ver-1.0 \citep[also see][]{dewangan12,dewangan13}. 
The aperture photometry was also performed on the MIPSGAL 24 $\mu$m image 
using a $7\arcsec$ aperture radius and a sky annulus from $7\arcsec$ to $13\arcsec$ in IRAF. 
MIPS zero-magnitude flux density, including aperture correction was taken from the MIPS Instrument Handbook-Ver-3. 

To examine the shock excited molecular hydrogen (H$_{2}$) in the region, we used a narrow-band H$_{2}$ v= 1-0 S(1) 2.12 $\mu$m image.
The v= 1-0 S(1) line of H$_{2}$ at 2.12 $\mu$m is considered as an excellent tracer of shocked regions. 
A continuum-subtracted narrow-band H$_{2}$ image was retrieved from the UKIRT Wide-field Infrared Survey for H2 (UWISH2) database
\citep{froebrich11}. The UWISH2 is an unbiased survey of the inner Galactic plane in the H$_{2}$ 1-0 S(1) line 
at 2.122 $\mu$m ($\Delta\lambda=0.021$ $\mu$m; a velocity resolution of about 2970 km s$^{-1}$) using WFCAM at UKIRT. 

To trace cold and dense region, we used dust continuum FIR and sub-millimeter emission maps. 
The continuum maps were obtained at 70 $\mu$m, 160 $\mu$m, 250 $\mu$m, 350 $\mu$m, and 500 $\mu$m 
as part of the {\it Herschel} Infrared Galactic Plane Survey \citep[Hi-GAL,][]{molinari10}.
The angular resolutions (beam FWHM) of these bands are 5$\farcs$8, 12$\arcsec$, 18$\arcsec$, 25$\arcsec$, and 37$\arcsec$, respectively. 
We downloaded the processed ``Standalone Browse Products'', using the {\it Herschel} Interactive Processing 
Environment \citep[HIPE,][]{ott10}. 
The aperture photometry for point sources was carried out in each of the {\it Herschel} maps using 
the Graphical Astronomy and Image Analysis (GAIA) tool. 
The radius of the circular apertures was $12\arcsec$, $22\arcsec$, $22\arcsec$, 
$30\arcsec$, and $42\arcsec$ at 70, 160, 250, 350, and 500 $\mu$m maps, respectively.
The background was estimated in a sky annulus from $20\arcsec$ to $30\arcsec$ and from $30\arcsec$ to $40\arcsec$ at 
70 and 160 $\mu$m, respectively, and in a sky annulus from $60\arcsec$ to $90\arcsec$ at 250, 350, and 500 $\mu$m. 
Aperture corrections were applied to the total fluxes in each {\it Herschel} band, 
which were taken from the {\it Herschel} Handbooks, available from the {\it Herschel} Science Centre website\footnote[2]{http://herschel.esac.esa.int/}. 
The sub-millimeter continuum map at 870 $\mu$m (beam size $\sim$ 19$\farcs$2) was also retrieved from the ATLASGAL 
archival survey\footnote[3]{http://www3.mpifr-bonn.mpg.de/div/atlasgal/}.  

In order to trace the ionized gas in the region, we used a radio continuum map at 20 cm, which we retrieved from 
the VLA Multi-Array Galactic Plane Imaging Survey \citep[MAGPIS;][]{helfand06} archive. 
These observations were carried out with the VLA configurations: B, C, and D. 
To correct for missing flux on large angular scales, Effelsberg 100-m data were combined with the VLA images.
Final processed images were made available with a 6$\farcs$2 $\times$ 5$\farcs$4 restoring beam 
and a pixel scale of $2\arcsec$/pixel. 

The observed positions of the 6 cm detections \citep{becker94}, EGOs \citep{cyganowski08}, 
ATLASGAL clumps \citep{contreras13}, 6.7 GHz methanol masers \citep{walsh98}, and water maser \citep{brand94} were 
collected from the literature. The IRAS flux densities were also obtained for the IRAS 18060-2005 source, using the IRAS point-source catalog \citep{helou88}.
The flux densities were found to be 140, 1040, 6030, and 12000 Jy at 12, 25, 60, and 100 $\mu$m, respectively. 

In addition, a published map at 21 cm \citep{kim01} and 
integrated molecular $^{13}$CO(2-1) line data \citep{beuther11} were also utilized in the present work. 
\section{Results}
\label{sec:data}
\subsection{A comparison of various signposts}
\label{subsec:evn}

\begin{figure*}
\includegraphics[width=\textwidth]{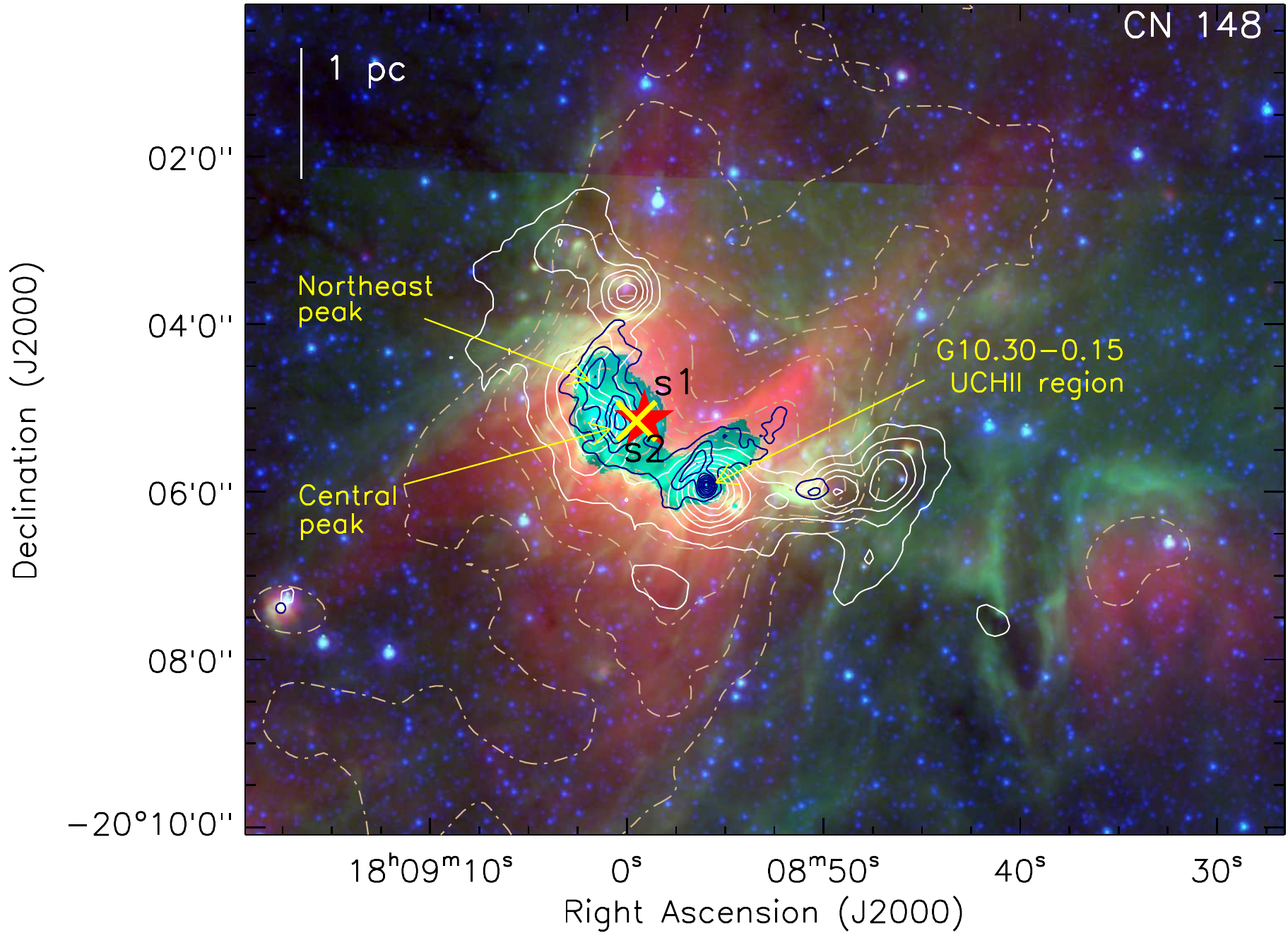}
\caption{Three color composite image of the bubble CN 148 in the G10.3-0.1 region (size of the selected field $\sim 12\farcm4  \times 9\farcm9$; central
coordinates: $\alpha_{2000}$ = 18$^{h}$08$^{m}$53$^{s}$.0, $\delta_{2000}$ = -22$^{\degr}$05$^{\arcmin}$08\farcs6) 
from {\it Spitzer} observations with MIPSGAL 24 $\mu$m (red), GLIMPSE 8 $\mu$m (green), and 4.5 $\mu$m (blue). 
ATLASGAL 870 $\mu$m emissions are overlaid by white contours with the levels of 0.76, 1.51, 2.27, 3.03, 4.17, 5.30, 
6.44, and 7.19 Jy/beam. New MAGPIS 20 cm radio continuum (angular resolution $\sim$ 6$\arcsec$) contours in navy blue color are 
also overlaid with 10, 20, 30, 40, 55, 70, 85, and 95\% of the peak value i.e. 0.206 Jy/beam. Tan color dot-dash contours 
delineate the previously published 21 cm radio 
continuum emission \citep[spatial resolution $\sim$ $37\arcsec \times 25\arcsec$; from][]{kim01} 
with 1, 3, 5, 10, 20, 30, 40, 55, 70, 85, and 95\% of the peak value i.e. 1.425 Jy/beam. 
The locations of an UCH\,{\sc ii} region G10.30-0.15 and two compact components are also shown in the figure (see Table~\ref{tab1}). 
Two candidates, s1 (18060nr1733) and s2 (18060nr2481), identified 
as O-stars (see the text) are also marked by big filled ``$\star$ (red)" and ``$\times$ (yellow)" symbols in the figure, 
respectively. The scale bar on the top left shows a size of 1 pc at a distance of 2.2 kpc.}
\label{fig2u}
\end{figure*}

\begin{figure*}
\includegraphics[width=\textwidth]{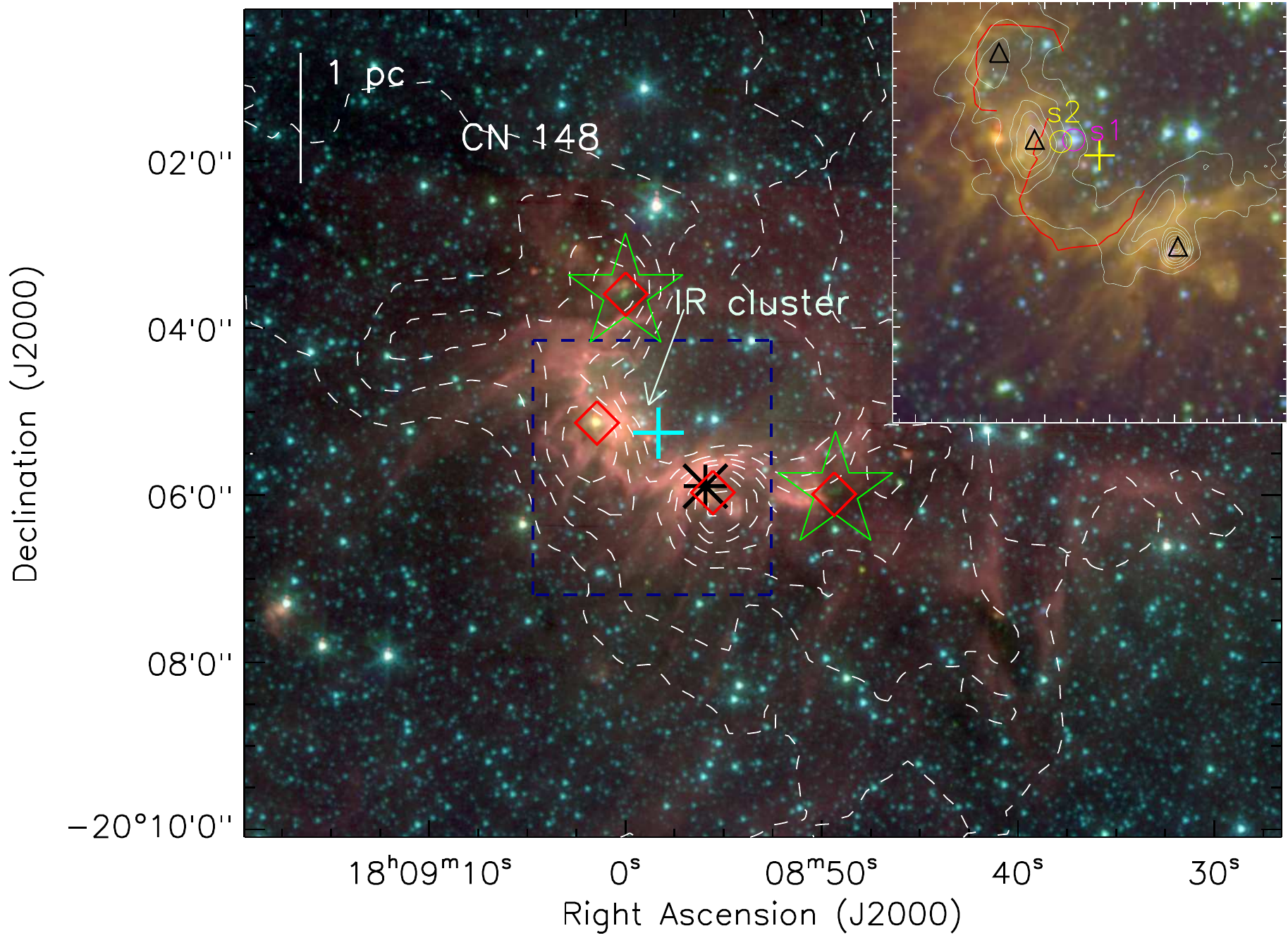}
\caption{{\it Spitzer}-IRAC 8.0 $\mu$m (red), 4.5 $\mu$m (green), and 3.6 $\mu$m (blue) images 
in log scale from {\it Spitzer}-GLIMPSE survey. 
The dashed contours are integrated $^{13}$CO(2-1) molecular gas emission from \citet{beuther11} 
with 10, 20, 30, 40, 55, 70, and 95\% of the peak value i.e. 153.26 K km s$^{-1}$.
The positions of IRAS 18060-2005 (+), four methanol masers ($\Diamond$), 
water maser (big asterisk) and two EGOs (big star symbols) are marked in the figure. 
The inset on the top right represents the central region in zoomed-in view, made of 
{\it Spitzer}-IRAC 5.8 $\mu$m (red), 3.6 $\mu$m (green) and UKIDSS {\it K} (blue) images 
(see the blue dashed box in the main figure). MAGPIS 20 cm radio continuum contours are also shown on the inset image with similar levels to those as 
shown in Figure~\ref{fig2u}. Two arc-like structures, three 6 cm radio detections ($\triangle$) and IRAS 18060-2005 (+) are also drawn 
in the inset image. Two O-type candidates (s1 (18060nr1733) and s2 (18060nr2481)) visible in the inset image are shown by circle symbols.}
\label{fig3u}
\end{figure*}

\begin{figure*}
\includegraphics[width=\textwidth]{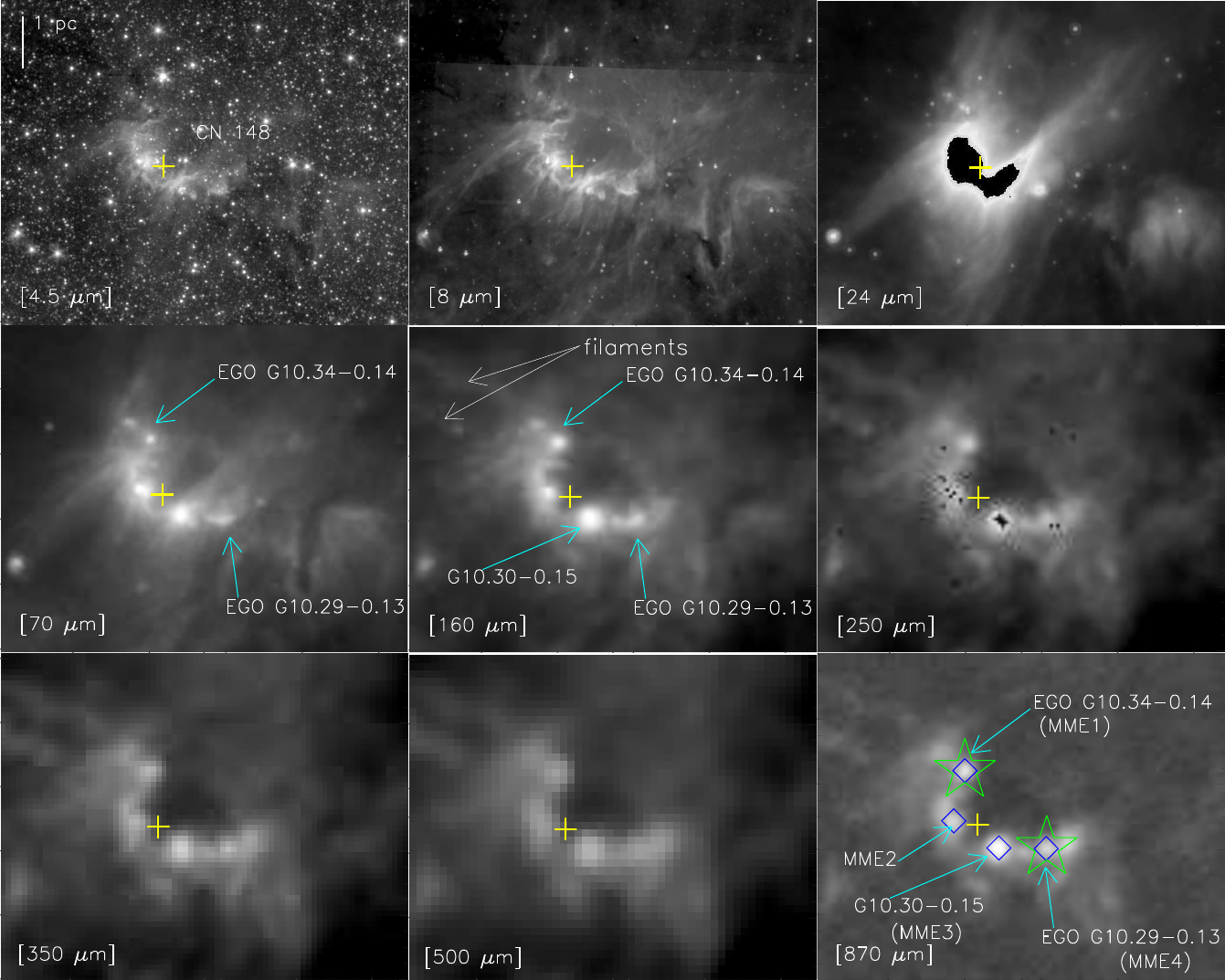}
\caption{A multi-wavelength view around the bubble CN 148.
The panels show images at 4.5 $\mu$m, 8.0 $\mu$m, 24 $\mu$m, 70 $\mu$m, 160 $\mu$m, 250 $\mu$m, 350 $\mu$m, 500 $\mu$m, 870 $\mu$m, 
from GLIMPSE, MIPSGAL, Hi-GAL, and ATLASGAL (from left to right in increasing order). 
The position of IRAS 18060-2005 is marked by a plus (+) symbol. The scale bar on the top left shows a size 
of 1 pc at a distance of 2.2 kpc. 
The 24 $\mu$m image is saturated near the IRAS position. 
Black dots seen at the 250 $\mu$m map can be an artifact. 
The 6.7 GHz methanol masers are also labeled as MME1-4 on 870 $\mu$m map. 
The other marked symbols are similar to those shown in Figure~\ref{fig3u} (see the text for more details).}
\label{fig4u}
\end{figure*}

Figure~\ref{fig1u} shows the W31 complex using Hi-GAL data. 
Different sub-regions of the complex are labeled in the figure. 
The location of CN 148 is denoted with a box. 
Figures~\ref{fig2u},~\ref{fig3u}, and~\ref{fig4u} provide an overview of the main characteristics of the bubble. 
Figure~\ref{fig2u} illustrates the spatial distribution of dust in the region, based on the 870 $\mu$m and 24 $\mu$m continuum 
data. The MAGPIS 20 cm radio continuum contours (blue color; angular resolution $\sim$ $6\arcsec$) are drawn. 
We note that the 24 $\mu$m emission is saturated in the proximity of the 20 cm continuum emission. 
We looked at the publicly available archival WISE\footnote[4]{Wide Field Infrared Survey Explorer, which is a joint project of the
University of California and the JPL, Caltech, funded by the NASA} \citep{wright10} image at 22 $\mu$m (spatial resolution $\sim 12\arcsec$) 
and found that it is equally saturated. 
The extended bipolar structure which is visible in the 21 cm data (see Introduction) appears to be traced also by the 
24 $\mu$m data (Figure~\ref{fig2u}). Both the warm dust and ionized emissions are well correlated, as pointed out 
by \citet{beuther11} for the extended H\,{\sc ii} region. 
In general, the 24 $\mu$m emission and ionized gas are found systematically 
correlated in H\,{\sc ii} regions \citep[e.g.][]{deharveng10,paladini12}. 
The MAGPIS 20 cm radio continuum map reveals finer features, and very weakly traces the extended emission (not seen in the figure). 
The outer (lower) radio contour shows an arc-like morphology whose convex surface faces the center of the bubble. 
The map at 20 cm traces an UCH\,{\sc ii} region (G10.30-0.15) and two additional compact H\,{\sc ii} regions, 
with sizes ranging from 0.22 to 0.27 pc (also highlighted in Figure~\ref{fig2u}). 
It is interesting to note that a small elongated ionized component (size $\sim 37\arcsec$ or 4.1 pc at a distance of 2.2 kpc) 
northeast of the UCH\,{\sc ii} region (separated approximately by $\sim$ 0.15 pc) is detected in the radio map. 
The elongation of the feature possibly indicates that its shape was influenced by the ionized gas flow. 
The position of two embedded sources, 18060nr1733/s1 (O5V-O6V) and 18060nr2481/s2 (O8V-B2.5V) \citep{bik05}, are also 
marked on the figure. 
The association between the MIR bubble, $^{13}$CO(2-1) molecular gas, maser positions, and the 
EGOs can be found in Figure~\ref{fig3u}. 
In this figure, both the MIR emission and the molecular gas 
display an almost circular morphology, except for the north and northwest directions, where the molecular density is low. 
The molecular gas emission is absent inside the bubble, and 
this empty emission region was previously reported as a molecular gas hollow by \citet{kim02}, which also measured its size, of the 
order of 2$\farcm$4 (1.5 pc at a distance of 2.2 kpc), which is very close 
to the size of the MIR bubble ($\sim$ 2$\farcm$8 or 1.8 pc) estimated by \citet{churchwell07} using the GLIMPSE 8.0 $\mu$m emission. 
\citet{becker94} observed three radio continuum sources in the region at 6 cm and with a 4$\arcsec$ angular resolution. Their 
peak positions are marked by triangles in Figure~\ref{fig3u}. 
These 6 cm peaks are coincident with the peaks of the MAGPIS 20 cm radio continuum emission, 
including the UCH\,{\sc ii} region G10.30-0.15. 
The positions of an UCH\,{\sc ii} region, four 6.7 GHz methanol masers, and a water maser are coincident 
at the edge of the bubble, indicating their physical association. 
Additionally, one can notice that there are two small arc-like structures prominently 
seen in all IRAC images (drawn in the inset in Figure~\ref{fig3u}), facing towards the center of the bubble. 
The K-band cluster (including O5V-O6V and O8V-B2.5V stars) is surrounded by one of the arc-like structures, 
and another arc-like structure harbors a radio emission peak, situated in the northeast direction. 
These two arc-like features demonstrate the impact of the ionizing photons. 
A {\it JHK} color composite image (not shown here) reveals the bright clustered regions (marked in Figure~\ref{fig3u}), 
along with the dark extincted patches of the molecular material. 

Figure~\ref{fig4u} presents a multi-wavelength view of the bubble CN 148.
The continuum data are shown at 4.5 $\mu$m, 8.0 $\mu$m, 24 $\mu$m, 70 $\mu$m, 160 $\mu$m, 250 $\mu$m, 350 $\mu$m, 500 $\mu$m, and 870 $\mu$m, 
from left to right in increasing order. The {\it Spitzer} images (4.5--24 $\mu$m) are shown only for comparison with the Hi-GAL maps. 
The Hi-GAL counterparts (70--500 $\mu$m) are mostly seen towards the four 
6.7 GHz methanol masers (including two EGOs) at the edges of the bubble. 
We labeled these methanol maser emissions MME1-4 (see Figure~\ref{fig4u}). 
The MIPS 24 $\mu$m image is saturated towards two 6.7 GHz masers (i.e. MME2 and MME3) and MME3 is associated with an UCH\,{\sc ii} region. 
The 24 $\mu$m emission is seen towards the two remaining 6.7 GHz methanol masers (i.e. MME1 and MME4), 
which are associated with the EGOs (i.e. G10.34-0.14 and G10.29-0.13). 
There is no discrete 24 $\mu$m counterpart found towards the EGO G10.29-0.13, while the EGO 
G10.34-0.14 is blended with other 24 $\mu$m source(s) \citep[also see][]{cyganowski09}. 
In particular, the MME4 in the southwest direction (towards EGO G10.29-0.13) appears 
as a compact emission rather than a diffuse source only at wavelengths longer than 
70 $\mu$m (see Figure~\ref{fig4u}). This suggests the presence of a deeply embedded discrete counterpart of the EGO G10.29-0.13/MME4, 
at the southwest edge of the bubble. Additionally, Hi-GAL data at wavelengths longer than 70 $\mu$m trace the filament-like 
features in the direction northeast from the bubble. 
The peaks of dust continuum emission are spatially coincident along the waist of the 
bipolar morphology, together forming the rim of the observed 
bubble. The north portion of the bubble is weakly detected, possibly because the so-called {\it bubble} could rather be a shell 
with a given orientation in the sky \citep[e.g.,][]{beaumont10}. 
Figures~\ref{fig2u},~\ref{fig3u}, and~\ref{fig4u} are prominently revealing only one rim of this shell. 
\subsubsection{Extinction maps}
\label{subsec:extinc}
The visual extinction (A$_{V}$) map of the region was generated using UKIDSS NIR (JHK) photometric data, 
following a similar procedure to the one described by \citet{kumar07}. The selection of the main sequence stars 
and the determination of the individual extinction values was performed using a NIR color-color diagram (J-H/H-K) 
with the reddening laws of \citet{rieke85}. The extinction map was created by Nyquist binning using the spatial distribution 
of the extinction values of individual stars, with a bin size of 60$\arcsec$. Following the suggestion of \citet{lombardi01}, 
we also performed sigma-clipping in our spatial smoothing, in order to minimize the effect of the foreground stars in the extinction map. 
These were identified as extremely low extinction outliers and removed from the local average. 
Figure~\ref{fig5u}a shows the extinction map (A$_{V}$ range 3-16 mag) of the region, overlaid with the $^{13}$CO(2-1) integrated intensity 
contours \citep{beuther11} and the positions of the dust clumps identified from the ATLASGAL survey at 
870 $\mu$m \citep{contreras13}. Within our selected region, we find five starless clumps from a 
catalog of \citet{tackenberg12} whose peak positions are also marked on Figure~\ref{fig5u}a. 
These clumps are located on the extreme outer boundary of ATLASGAL 870 $\mu$m emission, with NH$_{3}$ velocity of $\sim$ 13.6 km s$^{-1}$ and 
clump mass range of 120-420 M$_{\odot}$ \citep[see][for more detail]{tackenberg12}. 
All dense clumps are associated with white regions in the extinction map, indicating that the visual extinction value 
towards these clumps should be greater than 16 mag. The average extinction value of the region is found to be $\sim$ 7.8 mag. 
The extinction map traces the previously known K-band cluster with an A$_{V}$ of $\sim$ 14 mag near the IRAS position. 
All masers (MME1-4) appear to be located in the region having A$_{V}$ greater than 16 mag.
We suggest that these masers are not part of the K-band cluster, 
based on their spatial locations and relatively different extinction values with respect to the cluster. 
This interpretation is also supported by the fact that all stars in a cluster are assumed to have formed at about the same time.
\subsubsection{The PAH distribution and the H\,{\sc ii} region}
\label{subsec:ratmap}
\begin{figure*}
\includegraphics[width=14.25cm]{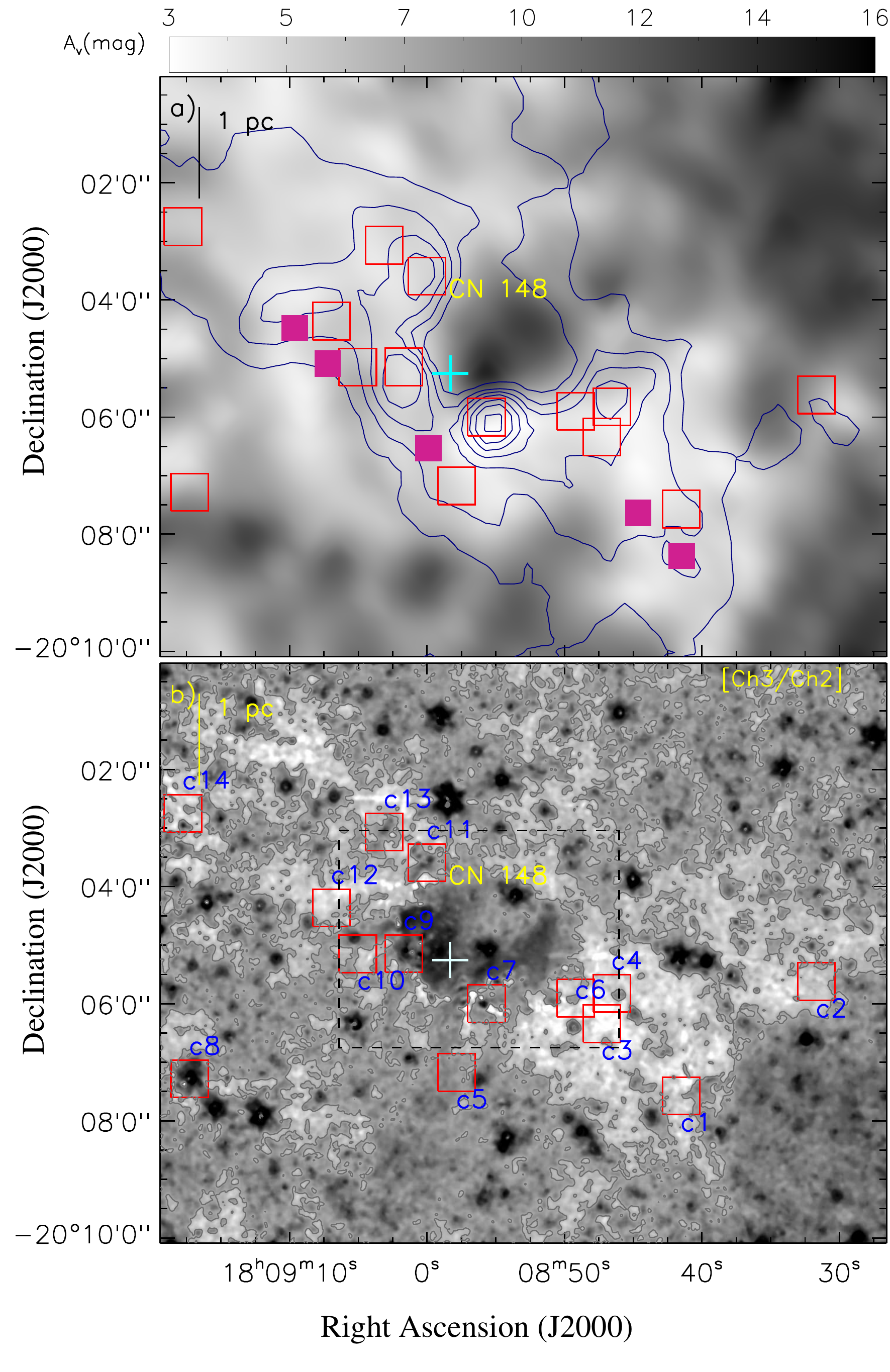}
\caption{a) Visual extinction (A$_{V}$) map of the region obtained using NIR data (see text for details). 
Integrated molecular $^{13}$CO(2-1) emission is overlaid with navy blue contours similar to those as 
shown in Figure~\ref{fig3u}. The peak positions of five starless clumps \citep{tackenberg12} are also marked by filled square symbols. 
b) {\it Spitzer}-IRAC ch3/ch2 ratio map around CN 148 (for a similar area as in Figure~\ref{fig2u}). 
The ch3/ch2 ratio contours are also shown on the image with a representative value of 8.3. 
The black dashed box is shown as a zoomed-in view in Figure~\ref{fig6u}. 
The positions of dust clumps are marked by big red square symbols in both images and labeled in Figure~\ref{fig5u}b (see Table~\ref{tab2}). 
Other marked symbols are the same as in Figure~\ref{fig3u}.}
\label{fig5u}
\end{figure*}

\begin{figure*}
\includegraphics[width=\textwidth]{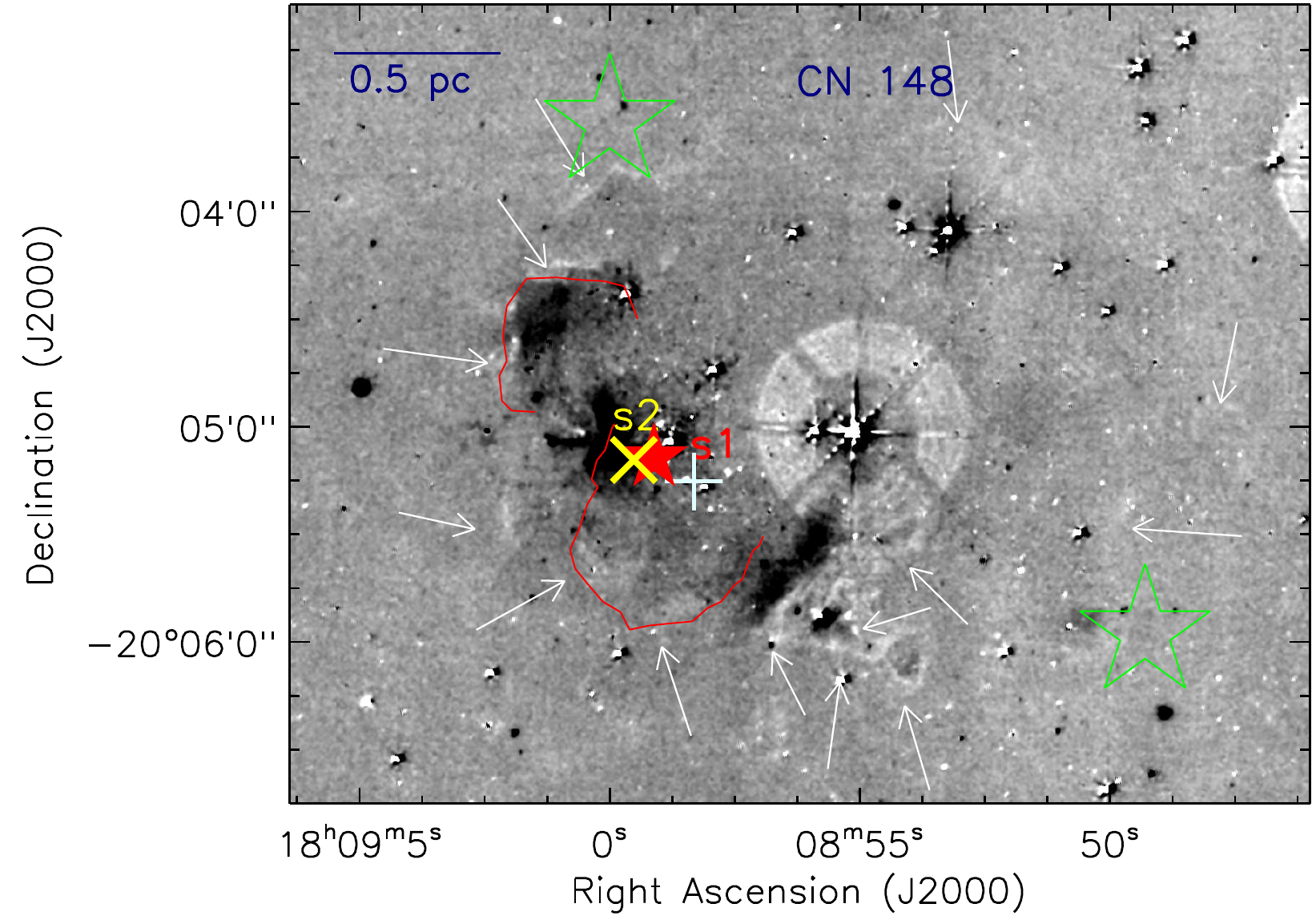}
\caption{Map of continuum-subtracted H$_{2}$ image (gray-scale) at 2.12 $\mu$m around CN 148 
(size of selected region in map $\sim$ 4.8 $\times$ 3.7 arcmin$^{2}$; as shown by a 
dashed box in Figure~\ref{fig5u}b). White arrows indicate the detected H$_{2}$ emission 
along the bubble (as seen in GLIMPSE images).
The marked symbols are similar to those shown in Figures~\ref{fig2u} and~\ref{fig3u}. 
Two red arc-like curves are similar as shown in Figure~\ref{fig3u}. 
The final continuum-subtracted H$_{2}$ image was obtained from 
UWISH2 survey \citep{froebrich11} and was further processed through median filtering with a 
width of 5 pixels and smoothened by 4 pix $\times$ 4 pix using a boxcar algorithm to trace 
the faint features in the image.}
\label{fig6u}
\end{figure*}

PAH features are known to trace photodissociation regions (or photon-dominated regions, or PDRs). 
PAH emission is produced in the PDR by ultraviolet (UV) photons ($6 eV < h \nu < 13.6 eV$) 
from massive stars exciting PAH molecules \citep{tielens04}. 
In our case, the presence of PAH emission allows us to infer the impact and influence 
of the expanding H\,{\sc ii} region on its surrounding. In recent years, ratio maps obtained using {\it Spitzer}-IRAC images (3.6--8.0 $\mu$m) have been utilized to trace 
PAH features at 3.3, 6.2, 7.7, and 8.6 $\mu$m, hence the extent of PDRs, in massive star-forming regions. 
In particular, the ch2 (4.5 $\mu$m) band can be used as a control band since it does not contain any known PAH line features. 
Therefore, IRAC ratio (ch4/ch2, ch3/ch2, and ch1/ch2) maps are generated using ch4 (8.0 $\mu$m), ch3 (5.8 $\mu$m), and ch1 (3.6 $\mu$m) 
bands with respect to the ch2 (4.5 $\mu$m) band. 
We generated ratio maps following a similar procedure as described by \citet{povich07} and \citet{dewangan11}. 
Figure~\ref{fig5u}b shows one such IRAC ratio map (ch3/ch2) overlaid with ratio contours. 
The bright features are indicative of PAH emission and allow us to trace the PDR surrounding the ionized gas. 
The ratio map also shows emission features similar to the filament-like emission seen in Hi-GAL images, in the northeast direction (see Figures~\ref{fig4u} and~\ref{fig5u}b). 
The positions of the ATLASGAL dust clumps are marked in Figure~\ref{fig5u}b. 
The comparison between the molecular CO, cold dust, and PAH emission is shown in Figures~\ref{fig5u}a and~\ref{fig5u}b. 
The molecular gas, cold dust, and PAH emission are distributed along the northeast to southwest directions (see Figures~\ref{fig5u}a and~\ref{fig5u}b). 
Figure~\ref{fig6u} shows a zoomed-in continuum-subtracted 2.12 $\mu$m H$_{2}$ image using the {\it K}-band image of 
the bubble. The figure reveals H$_{2}$ emission at the periphery of the source. 
The presence of H$_{2}$ emission can be explained by either UV fluorescence or shocks. 
A comparison of the line intensities of the 1-0 S(1) line of H$_{2}$ at 2.12 $\mu$m to the 2-1 S(1) line of H$_{2}$ at 2.25 $\mu$m is normally used to 
distinguish between fluorescent and shock-excited H$_{2}$ emission. 
However, this source was not observed in the 2-1 S(1) line of H$_{2}$ filter. 
Considering the observed morphology of the H$_{2}$ features, 
we suggest that the origin of the H$_{2}$ emission is likely caused by UV fluorescence.  
There are several arc-like structures of H$_{2}$ emission, mostly facing toward the center of the bubble, 
indicating the exposure to strong UV radiation. 
Figures~\ref{fig2u}--\ref{fig6u} suggest that the ionized region is dominant inside the bubble while the PAH, molecular gas, and cold dust emission 
are absent, as is expected, while the warm dust emission at 24 $\mu$m is detected in the  H\,{\sc ii} region, as mentioned before. 
Combining the multi-wavelength characteristics of CN 148, we suggest that the existence of the curved morphology of the molecular 
dense material is due to the expansion of the  H\,{\sc ii} region. 
Recently, \citet{arce11} predicted a circular or arc-like structure in the model for an expanding bubble \citep[also see][]{churchwell08,beaumont10},  
which further supports our argument on the observed molecular structure in CN 148. 
In summary, the molecular gas, PAH, cold dust, and H$_{2}$ emission 
are all coincident and constitute the evidence of star-forming material around the bubble. 
\subsubsection{Lyman continuum flux and ionizing feedback}
\label{subsec:lymnflx}
In this section, we derive the spectral type of the powering star(s) associated with the extended emission as well as the three components 
(UCH\,{\sc ii} region, central compact peak near the IRAS position, and northeast compact peak) detected in the radio maps.

We have calculated the number of Lyman continuum photons (N$_{uv}$) using the radio continuum map and 
following the formulation of \citet{matsakis76}:
\begin{equation}
\begin{split}
N_{uv} (s^{-1}) = 7.5\, \times\, 10^{46}\, \left(\frac{S_{\nu}}{Jy}\right)\left(\frac{D}{kpc}\right)^{2} 
\left(\frac{T_{e}}{10^{4}K}\right)^{-0.45} \\ \times\,\left(\frac{\nu}{GHz}\right)^{0.1}
\end{split}
\end{equation}
\noindent where T$_{e}$ is the electron temperature, D is the distance in kpc, $\nu$ is the frequency in GHz, and S$_{\nu}$ is the measured 
total flux density in Jy. S$_{\nu}$ was estimated for each of the individual components using the radio continuum map at 20 cm (1.4989 GHz).
Adopting D = 2.2 kpc and T$_{e}$ = 6800 K \citep[from][]{kim01}, we calculated N$_{uv}$. 
Table~\ref{tab1} summarizes the results. The Table also lists the spectral class of  each of the individual components 
inferred using the models of \citet[][his Table 2]{panagia73}. 
We find that each of the radio components is consistent with the radio spectral class of a B0V type star (see Table~\ref{tab1}). 
It is worth mentioning that the spectral class of potential ionizing sources is derived by 
assuming that all the ionizing flux is produced by only a single star. 

N$_{uv}$ is also estimated for the extended emission using 
S${_\nu}$ = 15.92 Jy at 21 cm (1.42 GHz) from \citet{kim01}, and is found to be $\sim$ 7.2 $\times$ 10$^{48}$ s$^{-1}$ (logN$_{uv}$ $\sim$
48.86). The MAGPIS map does not allow an accurate determination of the fluxes for extended sources ($>60\arcsec$), due 
to inadequate VLA u-v coverage \citep[see][for more details]{helfand06}. 
Therefore, the 21 cm data was preferred over MAGPIS 20 cm radio map as it detects the extended 
bipolar morphology (see Figure~\ref{fig2u}). 
The ionizing photon flux value corresponds to a single ionizing star of spectral type O7V-O6.5V \citep{panagia73,martins05}. 
It is obvious that the three B0V type stars together cannot produce sufficient ionizing radiation to explain the entire radio emission. 
Therefore, this extended emission is likely generated by other, more evolved sources located inside the bubble. 
It is to be noted that the two embedded sources 18060nr1733 (O5V-O6V) and 18060nr2481 (O8V-B2.5V) 
are associated with the K-band cluster located inside the bubble. The comparison of the spectral type of these two embedded sources with 
the radio spectral type of the powering source associated with the extended emission suggests 
that the spectroscopically identified O type sources can be ionizing candidates for the extended radio emission.

The radio continuum map at 20 cm is also used to estimate the 
electron density ($n_e$) of the ionized gas. 
The formula for ``$n_e$" is given by \citet{panagia78} under the assumption that the H\,{\sc ii} region
has a roughly spherical geometry:
\begin{equation}
\begin{split}
n_{e} (cm^{-3})=3.113\times 10^2 \left(\frac{S_\nu}{\rm{Jy}}\right)^{0.5}\left(\frac{T_e}{{\rm 10^4 K}}\right)^{0.25} 
\left(\frac{D}{{\rm kpc}}\right)^{-0.5} \\ \times\, b(\nu,T)^{-0.5}\theta_R^{-1.5}
\end{split}
\end{equation} 
In the equation above, S$_\nu$, $T_e$, and D are defined as in Equation 1, $\theta_R$ is the angular radius in arcminutes and 
\[ b(\nu, T)=1+0.3195~{\rm log}\,\left(\frac{T_e}{\rm 10^4 K}\right)-0.2130~{\rm log}\left(\frac{\nu}{\rm 1 GHz}\right).
\] The values of S$_\nu$ and $\theta_R$ used for each of the individual components and the 
corresponding values of $n_e$ are listed in Table~\ref{tab1}. The calculation was performed for D = 2.2 kpc and T$_{e}$ = 6800 K.

We have also estimated the internal pressure of a uniform density H\,{\sc ii} region (P$_{HII}$) to infer the ionizing feedback 
of massive stars on their environment. 
In Figures~\ref{fig2u} and~\ref{fig3u}, we have already noticed that the ionizing sources appear to be located inside the bubble. 
The molecular gas and dust emissions are absent inside the bubble and are seen towards the edges of the bubble. 
The pressure of an H\,{\sc ii} region is estimated at D$_{s}$ = 0.9 pc from the location of the O type stars, a distance similar to the radius of the bubble. 
The formula is given by $P_{HII}= \mu m_{H} c_{s}^2\, \left(\sqrt{3N_{uv}\over 4\pi\,\alpha_{B}\, D_{s}^3}\right)$; where, N$_{uv}$ is the number of 
ionizing photons emerging per second from the massive star, $\mu$ = 2.37, c$_{s}$ is the sound speed of 
the photo-ionized gas (= 10 km s$^{-1}$), and $\alpha_{B}$ is the radiative recombination coefficient \citep[=  2.6 $\times$ 10$^{-13}$ $\times$ (10$^{4}$ K/T$_{e}$)$^{0.7}$ cm$^{3}$ s$^{-1}$; see][]{kwan97}. 
One can find more details about this expression in the work of \citet{bressert12}.
Using N$_{uv}$ = 7.2 $\times$ 10$^{48}$ s$^{-1}$ and $\alpha_{B}$ = 3.4 $\times$ 10$^{-13}$ cm$^{3}$ s$^{-1}$ at T$_{e}$ = 6800 K, 
we calculate P$_{HII}$ $\approx$ 1.9 $\times$ 10$^{-9}$ dynes\, cm$^{-2}$.

In general, typical cool molecular clouds (particle density $\sim$ 10$^{3}$-10$^{4}$ cm$^{-3}$ and temperature $\sim$ 20 K) 
have pressure values $\sim$ 10$^{-11}$ -- 10$^{-12}$ dynes cm$^{-2}$ \citep[see Table 7.3 of][]{dyson97}. 
However, the pressure exerted by the self-gravity of the surrounding molecular gas in our 
selected region can be different than the pressure in a cool molecular cloud.
The pressure exerted by the self-gravity of the surrounding molecular gas is given by 
$P_{cloud}\approx\pi G\Sigma^2$; where $\Sigma$ $(= M_{cloud}/\pi R_{c}^2)$ is the mean mass surface density of the cloud, 
M$_{cloud}$ is the mass of the molecular gas, and R$_{c}$ is the radius of the molecular region. 
To estimate the total dust mass of the cloud including the bubble, 
we utilized the ATLASGAL 870 $\mu$m map and computed its total flux over the entire cold dust emission, as seen in Figure~\ref{fig2u}.
The integrated 870 $\mu$m flux is converted to the total dust mass of the cloud incorporating the gas-to-dust 
mass ratio (assumed to be 100), using a similar formula and inputs as mentioned in Section~\ref{sec:msdust}.
The total dust mass of the cloud is found to be $\sim$ 5490 $M_\odot$. 

Adopting $M_{cloud}$ $\approx$ 5490 $M_\odot$ (assuming about 10\% error in this value estimate) in a 
region of radius R$_{c}$ $\approx$ 3$\farcm$0 (or 1.9$\pm$0.3 pc at a distance of 2.2$\pm$0.3 kpc), 
we find a surface density $\Sigma \approx$ 0.10$\pm$0.03 g cm$^{-2}$ and 
$P_{cloud}$ $\approx$ 2.1$\pm$1.4 $\times$ 10$^{-9}$ dynes cm$^{-2}$, which is comparable to the value for P$_{HII}$ that we obtained earlier.  

We can also estimate the pressure contributions from the radiation pressure (P$_{rad}$ = $L_{bol}/ 4\pi c D_{s}^2$) and the stellar 
wind ram pressure (P$_{wind}$ = $\dot{M}_{w} V_{w} / 4 \pi D_{s}^2$). 
However, we do not have any specific information about the mass-loss rate ($\dot{M}_{w}$) 
and the wind velocity (V$_{w}$) of the powering star(s), and the bolometric luminosity (L$_{bol}$) of the region. 
The total FIR luminosity of IRAS 18060-2005 derived using the IRAS fluxes \citep[see][]{casoli86} is 
$\sim$ 0.94 $\times$ 10$^{5} L_{\odot}$ at a distance of 2.2 kpc. 
In order to derive P$_{wind}$ and P$_{rad}$ in the region, 
we have substituted the values $\dot{M}_{w}$ = 3.74 $\times$ 10$^{-7}$ M$_{\odot}$ yr$^{-1}$ 
\citep[for O6.5V star;][]{vink01}, V$_{w}$ = 2200 km s$^{-1}$ \citep[for O6.5V star;][]{kudritzki00}, and 
L$_{bol}$ $\approx$ 10$^{5} L_{\odot}$ \citep[see][]{beuther11,lumsden13}, in the above equations.
We find P$_{wind}$ $\approx$ 5.4 $\times$ 10$^{-11}$  dynes\, cm$^{-2}$ and $P_{rad}$ $\approx$ 1.3 $\times$ 10$^{-10}$ dynes\, cm$^{-2}$ 
at D$_{s}$ = 0.9 pc from the location of the O type sources. 
This calculation reveals that, in this case, the pressure of the H\,{\sc ii} region dominates over the 
radiation pressure and the stellar wind pressure, even assuming a factor of $10$ uncertainty associated with the estimated values. 
Summing up all the pressure components (i.e. P$_{HII}$, $P_{rad}$, and P$_{wind}$), we find that the total pressure, P$_{total}$, is 
$\approx$ 2.1 $\times$ 10$^{-9}$  dynes\, cm$^{-2}$. 
Therefore, the total pressure is in equilibrium with the pressure exerted by the surrounding, self-gravitating molecular 
cloud ($\approx$ 10$^{-9}$ dynes cm$^{-2}$), and suggests that the surrounding molecular cloud has been 
compressed to increase the pressure and balance the total pressure (P$_{total}$). This, in turn, could trigger the initial collapse 
and fragmentation of a large molecular cloud. 
Such a scenario is supported by the presence of several dense clumps surrounding the H\,{\sc ii} region. 
Altogether, the pressure of the H\,{\sc ii} region could be responsible for the expansion of the bubble and have 
significant influence on the surrounding environment. 
\begin{table*}
\setlength{\tabcolsep}{0.05in}
\centering
\caption{Physical parameters of the ionized components associated with the H\,{\sc ii} region, G10.3-0.1, using a 20 cm high resolution continuum map (see Figure~\ref{fig3u}).}
\label{tab1}
\begin{tabular}{lcccccc||cccccccc}
\hline 
  Source                 & RA         &  Dec       &  Angular radius		   & Total flux 	&  logN$_{uv}$   & Spectral Type       & Electron density  \\	
                         & [2000]     &  [2000]    & ($\theta_R$ in arcminutes)    &  (S${_\nu}$ in Jy) &   (s$^{-1}$)   &		      &  ($n_e$ in cm$^{-3}$)	   \\	
 \hline 		        		   			  	 								      
  UCH\,{\sc ii} region 	 & 18:08:55.9 & -20:05:54  &	0.36			   & 1.008		& 47.66 	 &    B0V--O9.5V   & 936.93 \\
  Central compact peak   & 18:09:00.4 & -20:05:10  &	0.42			   & 1.382		& 47.80 	 &    B0V--O9.5V   & 862.83   \\
  Northeast compact peak & 18:09:01.6 & -20:04:38  &	0.34			   & 0.724		& 47.52 	 &    B0.5V--B0V   & 850.42   \\
\hline          
\end{tabular}
\end{table*}
\subsubsection{Dust clumps}
\label{sec:msdust} 
In Section~\ref{subsec:extinc} we showed the spatial distribution of dust emission in CN 148 using the ATLASGAL 870 $\mu$m data. 
\citet{contreras13} describe the creation of the ATLASGAL compact source catalog, based on the source extraction algorithm SExtractor and using a 
detection threshold of 6-sigma. Noteworthy, the ATLASGAL catalog does not provide masses for the individual sources.
In this Section, we provide an estimate of the mass of the ATLASGAL sources in our selected region.
Within our selected region, we find 14 sources from the ATLASGAL catalog.
For a distance of 2.2 kpc, all the compact sources are clumps smaller than 1 pc (see Table~\ref{tab2}). 
For a reliable mass estimate, the observational knowledge about the temperatures within the clumps is crucial. 
\citet{wienen12} derived kinetic and rotational temperatures from NH$_{3}$ observations of the ATLASGAL sources. 
For our selected region, this information is available for only six clumps. 
The clump mass was estimated using the following formula \citep{hildebrand83}:
\begin{equation}
M \, = \, \frac{D^2 \, S_\nu \, R_t}{B_\nu(T_D) \, \kappa_\nu}
\end{equation} 
\noindent where $S_\nu$ is the integrated 870\,$\mu$m flux (Jy), 
$D$ is the distance (kpc), $R_t$ is the gas-to-dust mass ratio (assumed to be 100), 
$B_\nu$ is the Planck function for a dust temperature $T_D$, 
and $\kappa_\nu$ is the dust absorption coefficient. 
The clump masses and their associated errors are listed in Table~\ref{tab2}, assuming $\kappa_\nu$ = 1.85\,cm$^2$\,g$^{-1}$ \citep{schuller09},  
$D$ = 2.2$\pm$0.3 kpc and $T_D$ = 20 K (average rotational temperature of six ATLASGAL clumps; see Table~\ref{tab2}). 
Table~\ref{tab2} also includes the integrated 870\,$\mu$m fluxes (with uncertainties), clump effective radius, and NH$_{3}$(3,3) velocities, linewidths, 
rotational temperature and kinetic temperature. 
The total mass of 14 clumps is found to be $\sim$ 7264 M$_{\odot}$ and clump masses vary between 110 and 1533 M$_{\odot}$. 
Interestingly, four ATLASGAL clumps are physically associated with the 6.7 GHz methanol maser, and two of them 
are also physically associated with the EGOs. These clumps show high kinetic temperature as derived from the NH$_{3}$ line parameters (see Table~\ref{tab2}). 

In a previous study of the W31 region, \citet{beuther11} found several 870 $\mu$m cold dust emission clumps in the ATLASGAL survey, 
and estimated their masses using an average dust temperature of 50 K and a distance of 6 kpc. 
These authors used the CLUMPFIND algorithm at 3-sigma contour levels (210 mJy/beam) to extract the clumps from 870 $\mu$m map. 
They also adopted the dust absorption coefficient $\kappa_{870} \approx$ 0.8\,cm$^2$\,g$^{-1}$ and a gas-to-dust mass ratio of 186, in their calculation. 
Within our selected region, \citet{beuther11} identified 20 dust clumps with the total mass of $\sim$ 46440 M$_{\odot}$ 
with a mass range between 168 and 8187 M$_{\odot}$, at a distance of 6 kpc. 
One can notice the difference in number of clumps identified between the work of \citet{beuther11} and \citet{contreras13}, probably 
due to the selection of different clump extraction algorithms and different threshold values. A large difference in clump masses 
is also caused by the different choice of distance, dust temperature, dust absorption coefficient and gas-to-dust mass ratio in the calculation. 

The calculation of dust clump masses in this work therefore benefits from a better knowledge of the distance to the source and 
the availability of the NH$_{3}$ line parameters. 
\begin{table*}
\setlength{\tabcolsep}{0.05in}
\centering
\caption{Dust clumps from the ATLASGAL survey at 870 $\mu$m \citep{contreras13} in our selected region (also see Figure~\ref{fig5u}). 
Clump masses were estimated using integrated fluxes for a dust temperature = 20 K at a distance of 2.2$\pm$0.3 kpc.
NH$_{3}$ parameters (velocity (v(3,3)), linewidth (dv(3,3)), rotational temperature (Trot) and kinetic temperature (Tkin)) of 6 ATLASGAL clumps were 
taken from the literature for our selected region \citep[see][]{wienen12}. Superscript asterisk symbol (*) denotes the clumps 
associated with 6.7 GHz methanol masers. Association of the EGOs with the dust clumps are highlighted by superscript ``$\dagger$'' (see the text).}
\label{tab2}
\begin{tabular}{lcccccc||cccccccc}
\hline 
  ID        &  RA         &   Dec     & ATLASGAL         &  Effective      & Integrated 870 $\mu$m &Clump Mass     & v(3,3)   & dv(3,3)       & Trot     &  Tkin     \\   
            & [2000]      &  [2000]   & designation      &   Radius [pc]   &  Flux [Jy]            &  [M$_{\odot}$] &   [km/s]   &    [km/s]  &   [K]    &   [K]     \\   
\hline 						   	      	   							      
  c1 	                  & 18:08:41.5 & -20:07:34  & 010.249-00.111   &   0.83 &  20.37$\pm$3.34	      &   543.8$\pm$185.9	&    14.47  &	5.71  &   16.54  &   19.18   \\
  c2 	                  & 18:08:31.7 & -20:05:37  & 010.259-00.062   &   0.19 &   4.14$\pm$0.88	       &   110.5$\pm$40.6	&    ---    &	---   &   ---	 &   ---     \\
  c3 	                  & 18:08:47.3 & -20:06:20  & 010.278-00.121   &   0.69 &  16.64$\pm$2.78	      &   444.2$\pm$152.5	&    ---    &	---   &   ---	 &   ---     \\
  c4 	                  & 18:08:46.6 & -20:05:49  & 010.284-00.114   &   0.35 &  22.39$\pm$3.64	      &   597.7$\pm$203.9     &    ---    &	---   &   ---	 &   ---      \\
  c5 	                  & 18:08:57.8 & -20:07:10  & 010.286-00.164   &   0.44 &   9.95$\pm$1.76	      &   265.6$\pm$92.5	&    11.84  &	4.84  &   14.63  &   16.42    \\
  c6$^{*\dagger}$  & 18:08:49.2 & -20:05:54  & 010.288-00.124   &   0.16 &  10.74$\pm$1.88            &   286.7$\pm$99.6	&    ---    &	---   &   ---	 &   ---      \\
  c7$^*$                & 18:08:55.7 & -20:06:00  & 010.299-00.147   &   0.73 &  54.18$\pm$8.41	           &  1446.3$\pm$488.5	   &    12.69  &	6.75  &   22.98  &   30.20   \\
  c8 	                   & 18:09:17.3 & -20:07:17  & 010.321-00.231   &   0.35 &   5.21$\pm$1.04	      &   139.1$\pm$50.1	&    ---    &	---   &   ---	 &   ---     \\
  c9$^*$                & 18:09:01.7 & -20:05:08  & 010.323-00.161   &   0.91 &  57.42$\pm$8.90	            &  1532.8$\pm$517.6    &    12.41  &	3.45  &   24.05  &   32.36   \\
  c10	                  & 18:09:05.0 & -20:05:08  & 010.329-00.172   &   0.33 &  10.83$\pm$1.90	      &   289.1$\pm$100.5      &    ---    &	---   &   ---	 &   ---     \\
  c11$^{*\dagger}$ & 18:09:00.0 & -20:03:35  & 010.342-00.142   &   0.35 &  17.58$\pm$2.91             &   469.3$\pm$160.8	&    12.34  &	4.07  &   22.99  &   30.23   \\
  c12	                  & 18:09:06.9 & -20:04:21  & 010.344-00.172   &   0.37 &  12.89$\pm$2.21	      &   344.1$\pm$118.9	&    ---    &	---   &   ---	 &   ---     \\
  c13	                  & 18:09:03.1 & -20:03:04  & 010.356-00.149   &   0.51 &  24.32$\pm$3.93	      &   649.2$\pm$221.2	&    11.23  &	2.61  &   17.50  &   20.64    \\
  c14	                  & 18:09:17.8 & -20:02:44  & 010.388-00.196   &   0.16 &   5.47$\pm$1.08	       &   146.0$\pm$52.4	&    ---    &	---   &   ---	 &   ---      \\
\hline          
\end{tabular}
\end{table*}
\subsection{The YSO content of CN 148}
In the following, we present the identification and selection of young stellar populations using UKIDSS and GLIMPSE data, to learn about the star 
formation process around the bubble CN 148.
\subsubsection{Identification of YSOs}
\label{subsec:phot1}
\begin{figure*}
\includegraphics[width=\textwidth]{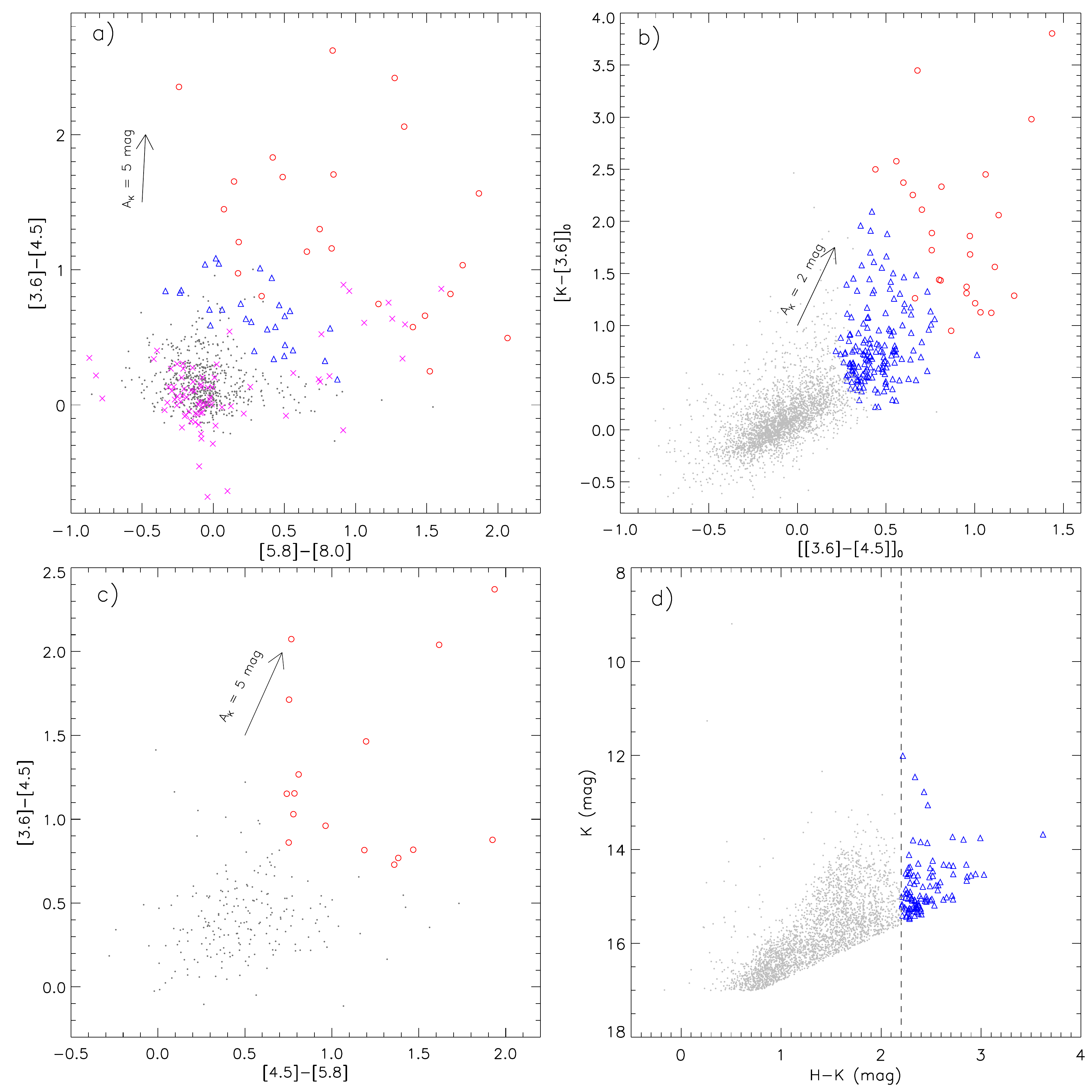}
\caption{Color-color and color-magnitude diagrams for all the sources identified within the region shown in Figure~\ref{fig2u}. 
a) Color-color diagram ([3.6]-[4.5] vs. [5.8]-[8.0]) using {\it Spitzer}-IRAC four band detections. 
The arrow shows the extinction vector with A$_{K}$ = 5 mag using the average extinction law from \citet{flaherty07}. 
The dots in gray color represent the stars with only photospheric emissions. 
The Class~0/I and Class~II YSOs are shown by open red circles and open blue triangles, respectively. 
The ``$\times$'' symbols in magenta color show the 
identified PAH-emission-contaminated apertures in the region (see the text). 
b) The dereddened [K - [3.6]]$_{0}$ $vs$ [[3.6] - [4.5]]$_{0}$ color-color diagram using NIR and GLIMPSE data. 
The arrow represents the extinction vector with A$_{K}$ = 2 mag using the average extinction law from \citet{flaherty07}. 
Open red circles and open blue triangles show Class I and Class II sources, respectively. 
c) Color-color diagram ([3.6]-[4.5] vs. [4.5]-[5.8]) of the sources detected in three IRAC bands, except 8.0 $\mu$m. 
The protostars are marked by open red circles (see the text for YSOs selection criteria). 
The arrow shows the extinction vector with A$_{K}$ = 5 mag using the average extinction law from \citet{flaherty07}. 
d) Color-magnitude diagram (K/H-K) of the sources detected only in H and K bands.
The YSOs are marked by open blue triangles (see the text for YSOs selection criteria).}
\label{fig7u}
\end{figure*}

The identification of YSOs is based on their IR excess emission, observed due to the presence of 
natal envelopes or circumstellar disks. 
The IR identification of these YSOs can be affected by various possible 
contaminants (e.g. broad-line active galactic nuclei (AGNs), PAH-emitting galaxies, shocked 
emission blobs/knots, and PAH-emission-contaminated apertures). 
\citet{gutermuth09} have proposed detailed criteria, using the four IRAC bands, to identify YSOs and remove the 
contaminants. It has been observationally reported that the interstellar extinction curve is relatively flat between 4.5 and 5.8 $\mu$m 
bands \citep{indebetouw05,flaherty07}. Therefore, the [4.5]-[5.8] color space is much less influenced by 
dust extinction and this color has been used for YSO classification by \citet{gutermuth09}. 
In order to identify YSOs and likely contaminants, we applied the various color conditions suggested by \citet{gutermuth09}.
The identified YSOs were further classified into different evolutionary stages (i.e. Class I, Class II, and Class III), 
using the slopes of the IRAC spectral energy distribution (SED) ($\alpha_{IRAC}$) measured from 3.6 to 8.0 $\mu$m 
\citep[e.g.,][]{lada06}. 
The $\alpha_{IRAC}$ conditions \citep[see][for more details]{billot10} were considered for 
Class I YSOs ($\alpha_{IRAC} > -0.3$), for Class II YSOs ($-0.3 > \alpha_{IRAC} > -1.6$), and for Class III sources ($-1.6 > \alpha_{IRAC} > -2.56$). 
The IRAC color-color diagram ([3.6]-[4.5] vs [5.8]-[8.0]) is shown in Figure~\ref{fig7u}a for all the identified sources. 
Using this approach, we obtain 50 YSOs (23 Class 0/I; 27 Class II), 630 photospheres, and 79 contaminants 
in the selected region around the bubble CN 148. 

Compared to the two shorter wavelengths (3.6 and 4.5 $\mu$m), the 5.8 and 8.0 $\mu$m IRAC bands are more sensitive to 
diffuse emission than point sources. 
Therefore, IRAC 3.6 and 4.5 $\mu$m bands were combined with UKIDSS NIR HK photometry (i.e. WFCAM-IRAC) to 
identify additional YSOs, mainly when sources were not detected in the IRAC 5.8 and/or 8.0 $\mu$m bands. 
We adopted the method developed by \citet{gutermuth09} to identify YSOs using WFCAM-IRAC (H, K, 3.6, and 4.5 $\mu$m) data \citep[see][for more details]{gutermuth09}.
In this scheme, the dereddened colors ([K - [3.6]]$_{0}$ and [[3.6] - [4.5]]$_{0}$) were calculated using the color 
excess ratios listed in \citet{flaherty07}. The classified Class I and Class II YSOs were further checked for possible dim contaminants by means of 
the dereddened 3.6 $\mu$m magnitudes (i.e.[3.6]$_{0}$ $<$ 14.5 for Class II and [3.6]$_{0}$ $<$ 15 for Class I), which were 
estimated using our extinction measurements and the reddening law from \citet{flaherty07}. 
Following this method, we found 191 additional YSOs (27 Class I and 164 Class II; see Figure~\ref{fig7u}b).

We also made use of an IRAC color-color diagram ([3.6]-[4.5] vs. [4.5]-[5.8]) to identify protostars 
when sources were either not detected or saturated in the 8.0 $\mu$m band. 
The protostars were selected using the criteria [3.6]-[4.5] $\ge$ 0.7 and [4.5]-[5.8] $\ge$ 0.7 from \citet{hartmann05} and \citet{getman07}. 
With this procedure, we identified 16 protostars (see Figure~\ref{fig7u}c).
 
There are a large number of sources detected only in the NIR regime, which are not present in the GLIMPSE data.
Therefore, we have selected very red sources having H-K $>$ 2.2 using the color-magnitude (K/H-K) 
diagram (see Figure~\ref{fig7u}d), and found 114 sources. 
This cut-off criterion allows us to identify very deeply embedded sources.
 
Combining the identification from all the four methods described above, a total of 371 YSOs are obtained in the region. 
The positions of all YSOs are plotted in Figure~\ref{fig8u}a.\\\\ 
The above methods can lead to contamination from other intrinsically ``red sources'' such as 
asymptotic giant branch (AGB) stars \citep{whitney08,robitaille08}. 
We checked the possibility of intrinsically ``red sources'' as well as AGB contaminations in our YSO sample, 
following the criteria suggested by \citet{robitaille08}, using 4.5 $\mu$m, 8.0 $\mu$m, and 24 $\mu$m photometry data. 
Only 50 out of the 371 YSOs have detections in the 4.5 $\mu$m and 8.0 $\mu$m bands. 
We find 21 out of these 50 YSOs ($\sim$ 42\%) to be possible intrinsically ``red sources''. 
These ``red sources'' were further checked for any likely AGB contaminations, 
based on their detections in the MIPSGAL 24 $\mu$m image. 
We find no AGB contamination among the selected ``red sources''. 
Additionally, AGB stars can also be disentangled from YSOs by applying a color-color criterion based on the 
Hi-GAL 70--350 $\mu$m flux densities \citep[e.g.,][]{veneziani13}. 
In our case, the application of the \citet{veneziani13}'s AGB identification method retrieved no AGB candidate.
\subsubsection{Spatial distribution of the YSOs and their clustering}
\label{subsec:surfden}
\begin{figure*}
\includegraphics[width=13.8cm]{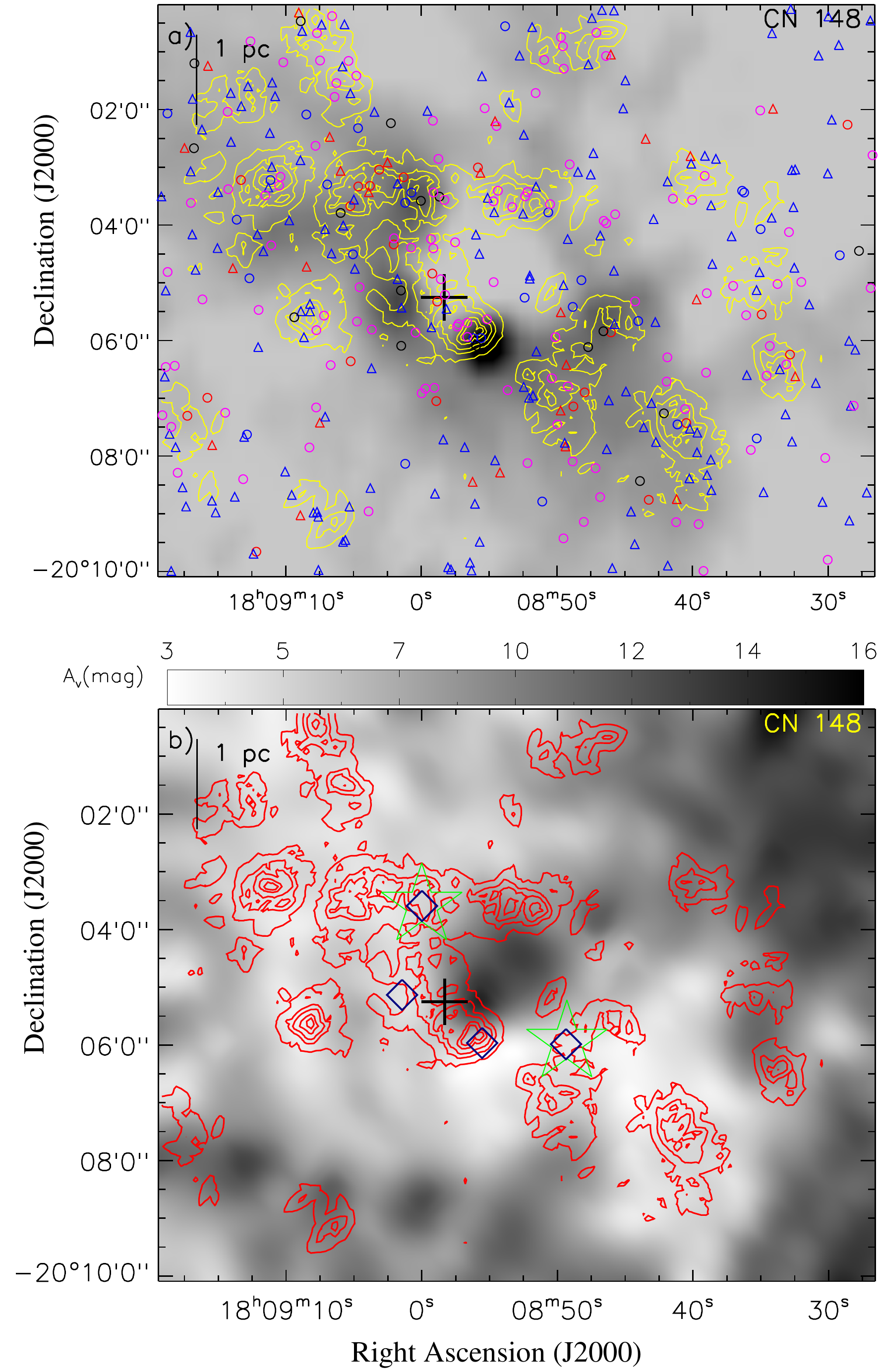}
\caption{a) The spatial distribution of YSOs and molecular $^{13}$CO(2-1) gas emission in the selected region around CN 148. 
The YSO surface density contours are drawn in yellow color for 10, 15, 20, 30, and 40 YSOs/pc$^{2}$, from the outer to the inner side (see text for details). 
The open circles and open triangles show the Class I and Class II sources, respectively. The YSOs identified using four IRAC bands, WFCAM-IRAC, 
three IRAC bands and red sources (H-K $>$ 2.2) are shown by red, black, blue, and magenta colors, respectively. 
The position of IRAS 18060-2005 (+) is also marked in the plot. 
b) The YSO surface density contours are drawn in red color similar levels to as shown in Figure~\ref{fig8u}a, 
and are superimposed on the visual extinction map. Other marked symbols are the same as in Figure~\ref{fig3u}.} 
\label{fig8u} 
\end{figure*}

In order to study the spatial distribution of the YSOs, we have constructed the YSO surface density map, following the method suggested 
by \citet{gutermuth09}. The surface number density at the {\it i$^{th}$} grid point is defined as $\rho_{i} = (n-1)/A_{i}$ with fractional uncertainty 
of (n -2)$^{-0.5}$ \citep{casertano85}, where $A_{i}$ is the surface area defined by the radial distance to 
the $n$ nearest-neighbour. We computed the 6$^{th}$ nearest-neighbour (NN) density map projected on 
a 5$\arcsec$ grid at a distance of 2.2 kpc. It is shown as contours on Figure~\ref{fig8u}. 
The choice of $n$ = 6 was found to be a good compromise between the resolution and sensitivity of the surface 
density calculation \citep[e.g.][]{casertano85,gutermuth09}. 
The contour levels are drawn at 10 YSOs/pc$^{2}$ (2.2 $\sigma$), 15 YSOs/pc$^{2}$ (3.2 $\sigma$), 20 YSOs/pc$^{2}$ (4.3 $\sigma$), 
30 YSOs/pc$^{2}$ (6.5 $\sigma$), and 40 YSOs/pc$^{2}$ (8.6 $\sigma$), increasing from the outer to the inner region. 
In Figure~\ref{fig8u}a, the overlay demonstrates the correlation of molecular gas emission with that of YSO surface density. 
The spatial distribution of the YSOs, superimposed on the visual extinction map, is shown in Figure~\ref{fig8u}b. 

\begin{figure*}
\includegraphics[width=15.70cm]{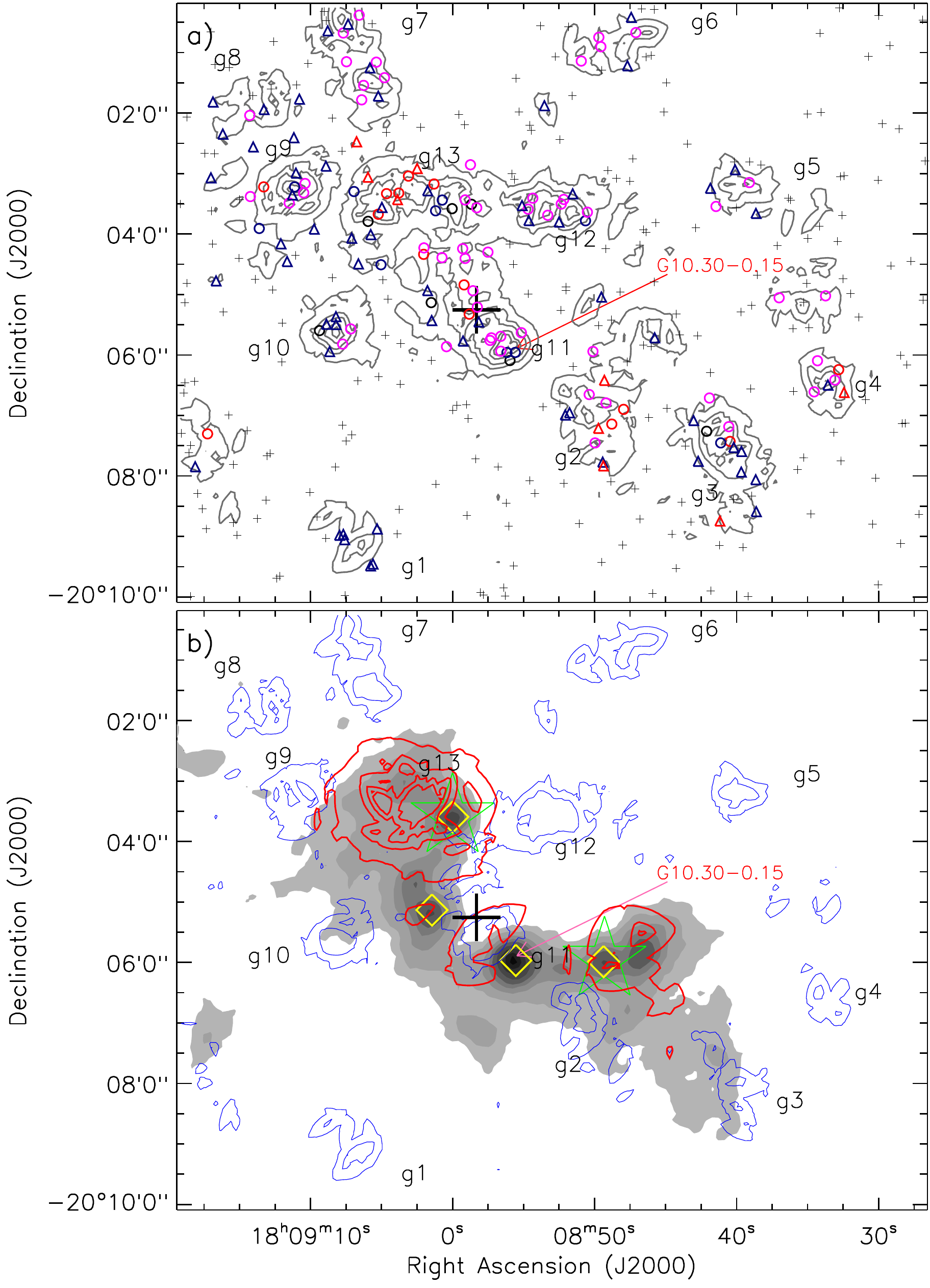}
\caption{a) All the selected cluster YSO members with d$_{c} \lesssim$ 0.47 pc are shown by open circles (Class I) and open triangles (Class II). 
The YSOs greater than d$_{c}$ (= 0.47 pc) are marked by plus symbols (black) (see text for details). 
The identified clusters are labeled in the diagram (see Table~\ref{tab3}). 
The other marked symbols are similar to those shown in Figure~\ref{fig8u}a. 
b) Comparison between the surface density contour maps of Class I and Class II YSOs 
in the selected region around the bubble CN 148. 
The surface density contours of Class I YSOs are drawn in red color with 3.5, 6, 8, and 10 YSOs/pc$^{2}$. 
The Class II YSO surface density contours are overlaid in blue color with 10 and 15 YSOs/pc$^{2}$. 
The background gray area indicates the contour map of ATLASGAL 870 $\mu$m emissions 
with 5, 10, 20, 30, 40, 55, 70, 85, and 95\% of the peak value i.e. 7.57 Jy/beam. 
The other marked symbols are similar to those shown in Figures~\ref{fig3u} and~\ref{fig9u}a.} 
\label{fig9u} 
\end{figure*}

To study the YSO clustering, we need to separate the clustered and scattered YSO populations.
This analysis can be done by determining the empirical cumulative distribution (ECD) of YSOs as a function of NN distance.
The ECD allows us to choose a cutoff length (also referred to as the distance of inflection d$_{c}$) that 
separates the low-density components \citep[see][for more details]{chavarria08,gutermuth09}. 
Using the ECD analysis, we find d$_{c}$ $\sim$ 0$^{\degr}$.0122 (or 0.47 pc at a distance of 2.2 kpc) 
to trace the cluster members within the contour level of 10 YSOs/pc$^{2}$ in the entire region. 
More details on the cluster identification can be found in the work of \citet{bressert10}. 
Considering the surface density level of 10 YSOs/pc$^{2}$, about 41\% of the YSOs (i.e. 154 from a total of 371 YSOs) are found in the clusters. 
We identified 13 clusters and label them in Figure~\ref{fig9u}a. 
This finding reveals the embedded clusters of YSOs (having extinction greater than $\sim$ 16 mag), 
mostly concentrated in the northeast to southwest, following the molecular gas emission around the bubble (see Figures~\ref{fig8u} and~\ref{fig9u}a). 
The spatial locations of the clusters clearly demonstrate that star formation is concentrated in 
the high extinction regions around CN 148.

All the clusters are listed in Table~\ref{tab3}, along with the number of members for each cluster. 
We notice age spread in the clusters with respect to the bubble, based on the relative distribution of Class I and Class II 
YSOs in each cluster. 
In order to trace an apparent age gradient in the clusters, we constructed the surface density maps of 
Class I and Class II YSOs (including sources having H-K $>$ 2.2) separately, following the procedure described above. 
The surface density maps of Class I and Class II YSOs are shown in Figure~\ref{fig9u}b. 
The surface density of Class I YSOs is found to vary between 0.16 and 17.11 YSOs/pc$^{2}$ with 
a dispersion($\sigma$) of 1.41 and it is drawn at contour levels of 3.5, 6, 8, and 10 YSOs/pc$^{2}$. 
The surface density of Class II YSOs varies between 0.67 and 40.68 YSOs/pc$^{2}$, with a standard deviation of 3.37 
and is drawn at contour levels of 10 and 15 YSOs/pc$^{2}$. 

Considering the observational evolutionary classification of YSOs, one can suggest that the clusters 
associated with Class I YSOs are relatively younger than those associated with Class II YSOs. 
The Class I cluster, ``g13", including an EGO (G10.34-0.14), is located at the northeast edge of the bubble. 
The Class II clusters, ``g7, g8, and g9", are spatially distributed in the northeast direction, 
further away from the Class I cluster ``g13". Therefore, in the northeast direction, there appears to be an apparent age gradient in the clusters with respect to the bubble. 
The ``g2 and g11" clusters contain both Class I and Class II YSOs (see Figures~\ref{fig9u}a and~\ref{fig9u}b). 
The cluster ``g11" contains an UCH\,{\sc ii}  G10.30-0.15 region, whereas the cluster ``g2" is associated with 
an EGO (G10.29-0.13) at the southwest edge of the bubble.  
\citet{cyganowski11}, based on VLA observations, found no 3.6 cm and 1.3 cm emission towards this EGO (G10.29-0.13). 
It is worth noting that both the UCH\,{\sc ii} and the EGO (G10.29-0.13) are associated with a 6.7 GHz methanol maser. 
\citet{chambers09} pointed out that the EGO represents the earliest phase of MSF 
before the onset of a hyper-compact(HC)/UCH\,{\sc ii} phase. 
In the later phase of evolution, an EGO phase ends with the formation of the HC/UCH\,{\sc ii} region.  
Due to these observational facts, we can suggest that the cluster ``g11" is relatively more evolved, compared to the cluster ``g2". 
The Class II clusters ``g3 and g4" are located at the southwest of cluster ``g2", 
suggesting an apparent age gradient in the clusters situated in the southwest 
direction with respect to the bubble. 

Overall, these observational characteristics indicate that the clusters at the edges of the bubble (both northeast and southwest) are relatively younger than the 
clusters located further away from the bubble. 
\begin{table*}
\scriptsize
\setlength{\tabcolsep}{0.05in}
\centering
\caption{The list of identified embedded clusters and their respective YSO populations. 
The YSO populations (Class I and Class II) are identified using four methods (i.e., four IRAC bands, three IRAC bands, 
WFCAM-IRAC (H, K, 3.6 $\mu$m and 4.5 $\mu$m) and H-K $>$ 2.2). 
All the clusters are labeled in Figure~\ref{fig9u}a (see the text for more details).}
\label{tab3}
\begin{tabular}{lcccccc||cccccccc}
\hline 
Cluster & \multicolumn{3}{c}{Class I}& \multicolumn{2}{c}{Class II} & \multicolumn{2}{c}{Total}& Red sources    \\
  ID    &  Four IRAC & WFCAM-IRAC & Three IRAC & Four IRAC &WFCAM-IRAC&Class I & Class II & H-K $>$ 2.2  \\
\hline 					   	      	              							      
  g1    & 0          &  0      &  0        & 0         & 4    &   0    & 4   &   0 \\
  g2    & 2          &  0      &  0        & 3         & 4    &   2    & 7   &   4 \\
  g3    & 1          &  1      &  1        & 1         & 7    &   3    & 8   &   2 \\
  g4    & 1          &  0      &  0        & 1         & 1    &   1    & 2   &   3  \\
  g5    & 0          &  0      &  0        & 0         & 2    &   0    & 2   &   2 \\
  g6    & 0          &  0      &  0        & 0         & 2    &   0    & 2   &   4 \\
  g7    & 0          &  0      &  0        & 0         & 4    &   0    & 4   &   7 \\
  g8    & 0          &  0      &  0        & 0         & 3    &   0    & 3   &   1 \\
  g9    & 1          &  1      &  0        & 0         & 3    &   2    & 3   &   3 \\
  g10   & 0          &  1      &  0        & 0         & 4    &   1    & 4   &   2 \\
  g11   & 0          &  2      &  1        & 0         & 2    &   3    & 2   &   5 \\
  g12   & 0          &  1      &  0        & 0         & 4    &   1    & 4   &   6 \\
  g13   & 5          &  3      &  2        & 3         & 2    &  10    & 5   &   1 \\
\hline          
\end{tabular}
\end{table*}
\subsubsection{Masers and the evolutionary stages of MSF}
\label{subsec:maser}
It was pointed out in the Introduction that the masers (four methanol and one water), an UCH\,{\sc ii} region, and two 
EGOs are physically associated with the bubble CN 148. 
It was also noted earlier that all the 6.7 GHz masers are associated with high-mass 870 $\mu$m dust 
continuum clumps (see Table~\ref{tab2}). 
It is interesting to recall that the masers MME1, MME3, and MME4 are associated with the EGO G10.34-0.14, EGO G10.29-0.13, and the UCH\,{\sc ii} region, respectively. 
Our photometric analysis reveals a Class~I YSO towards the maser MME2. The position of this Class~I YSO coincides with the maser 
position to within $2\arcsec$. \citet{moises11} also identified an embedded source towards this maser and 
suggested that the source is a probable MYSO candidate. It is found to be saturated in the 8 and 24 $\mu$m images and is 
coincident with the ATLASGAL clump ``c9'' (see Figure~\ref{fig5u}). 
The outcome of the SED modelling of the IR-counterpart (IRc) of MME2 also supports the argument 
that this source is a MYSO (see Section~\ref{subsec:sed}). 
Based on the results, we can suggest the presence of different early evolutionary stages of 
MSF (MYSO, EGOs, and a relatively evolved UCH\,{\sc ii} region) at the edges of the bubble. 
These stages do not appear to be associated with the K-band IR cluster located inside the bubble.
\subsubsection{The SED modelling}
\label{subsec:sed}
\begin{figure*}
\includegraphics[width=14.70cm]{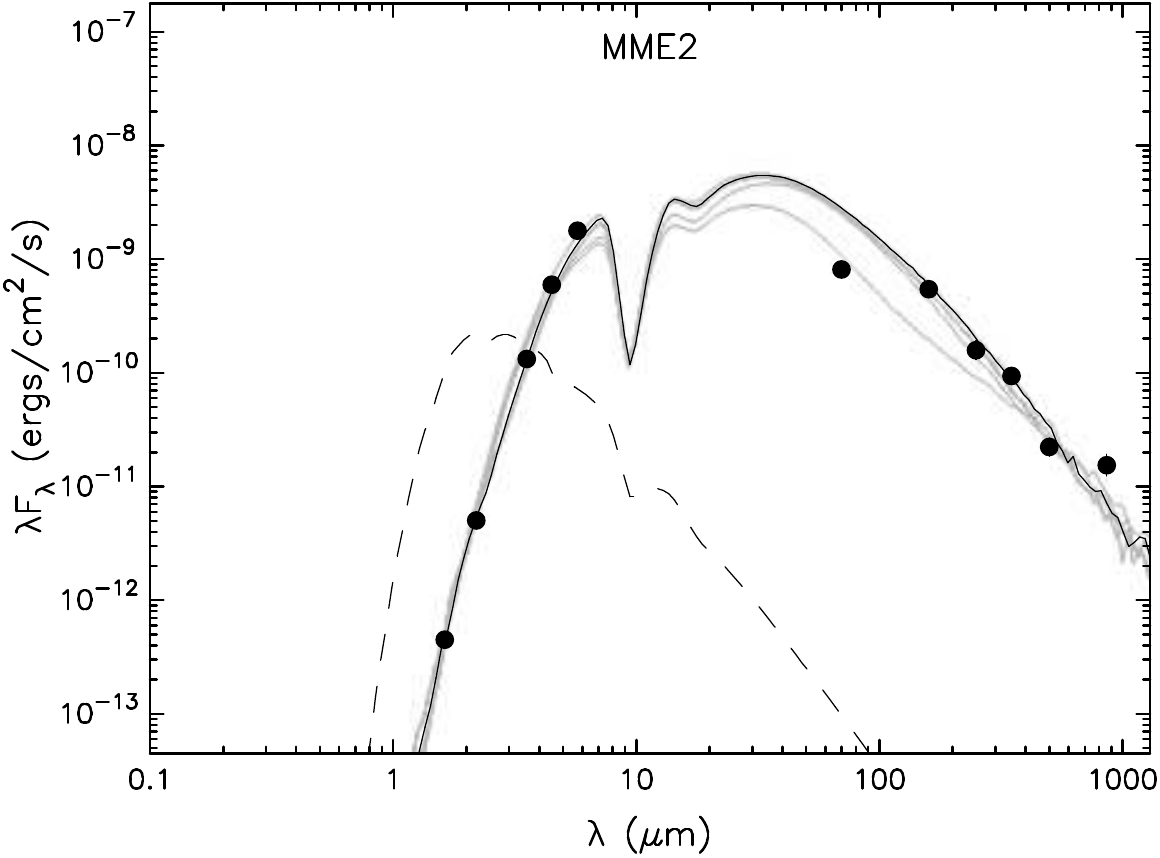}
\caption{The SED plot of the IRc of the maser MME2. 
The filled circles are observed fluxes of good quality (see Section~\ref{subsec:sed}) and the grey curves show the fitted model 
for ($\chi^{2}$ - $\chi^{2}_{best}$) per data point $<$ 5. The thin black curve corresponds to the best fitting model. 
The dashed curves represent photospheric contributions.} 
\label{fig10u} 
\end{figure*}

We modelled the SED of the IRc of MME2 ($\alpha_{2000}$ = 18$^{h}$09$^{m}$01.2$^{s}$, $\delta_{2000}$ = -20$\degr$05$\arcmin$09$\arcsec$) 
using an on-line SED modeling tool \citep{robitaille06,robitaille07}.
The model grid consists of 20,000 SED models from \citet{robitaille06} computed using two-dimensional radiative transfer 
Monte Carlo simulations. 
Each YSO model provides the output SEDs for 10 inclination angles, covering a range of stellar masses from 0.1 to 50 M$_{\odot}$.
These models assume an accretion scenario with a central source associated with 
rotationally flattened infalling envelope, bipolar cavities, and a flared accretion disk, all under radiative 
equilibrium. We only select those models which satisfy the criterion: $\chi^{2}$ - $\chi^{2}_{best}$ $<$ 5, 
where $\chi^{2}$ is taken per data point. The SED fitted models for the IRc of MME2 are shown in Figure~\ref{fig10u} . 
The IRc is detected at wavelengths longer than 1.25 $\mu$m. 
The weighted mean values of the luminosity and mass are 2355$\pm$940L$_{\odot}$ and 7.7$\pm$2.4M$_{\odot}$, respectively.
Our SED result suggests that the IRc of MME2 could be a MYSO. 
\section{Discussion}
\label{sec:data1}
Our present study expands upon the results of \citet{beuther11} and \citet{kim01,kim02} with a 
better distance estimate to CN 148 and a detailed analysis of its stellar content. 
The ATLASGAL cold dust clumps (c1--c14) and previously published CO data \citep{beuther11} show the fragmentation of 
dense material in the region (see Figure~\ref{fig5u}). 
Six of these cold dust clumps, detected in NH$_{3}$ emission, are located at the edge of the 
bubble (see Table~\ref{tab2}), representing a high density (critical density $\sim$ 10$^{4}$ cm$^{-3}$) component of the molecular cloud. 
These clumps are found to be massive with a total dust mass of $\sim$ 4907 M$_{\odot}$. 
The clumps associated with 6.7 GHz masers show higher rotational and kinetic temperatures, and also display large line widths of 
the NH$_{3}$(3,3) line, which are indirectly suggestive of increased star formation activity \citep[also see,][]{wienen12}. 
In addition, five starless clumps are also spatially located on the extreme outer 
boundary of ATLASGAL 870 $\mu$m emission, away from the inner edges of the bubble (see Figures~\ref{fig5u}a and~\ref{fig9u}). 
We also note that there is an absence of IR excess source(s) in these starless clumps. 
All these dust clumps, including the starless ones, were exclusively found in regions with A$_{V} > 16$ mag.

In this paper, we have investigated the population of young stars using NIR and MIR data, 
which was lacking from previous studies of this region. Based on the identification of YSOs, we studied their spatial distribution to explore 
the details of the star formation process in CN 148. 
The surface density of YSOs is well correlated with the star-forming material in the PDR, 
mostly concentrated in the northeast to southwest axis with respect to the bubble (see Figures~\ref{fig8u} and~\ref{fig9u}). 
We have observed that the clusters at the edges of bubble (in northeast and southwest directions) are relatively younger than the clusters located 
further away from the bubble (see Figure~\ref{fig9u}b). 

As mentioned earlier, the early evolutionary stages of MSF (MYSO, EGOs, and relatively evolved UCH\,{\sc ii} region) can be seen 
at the edges of the bubble associated with the 6.7 GHz methanol masers. 
\citet{beuther11} also pointed out three different evolutionary stages of star formation such as a more 
evolved IR cluster, high-mass protostellar objects, and 
high-mass starless clump candidates, and suggested their formation due to the triggering process. 
This study reveals that the bubble CN 148 is one of the rare single targets, which contains different early phases 
of MSF at the edges, including starless clumps as well as significant clustering of YSOs. 

Recently, \citet{kendrew12}, \citet{thompson12}, and \citet{deharveng10} studied a large number of MIR bubbles 
to understand the statistical importance of triggered star formation and suggested the formation of 14--30\% of 
massive stars in the Milky Way by the triggering process.
Including these recent statistics, it is immediately evident that some fraction of these MIR bubbles associated with H\,{\sc ii} 
regions can be sites of triggered star formation. 

Altogether, the existence of an apparent age gradient in YSO clusters, their spatial distribution in the PDR, and 
different early evolutionary stages of MSF around the bubble powered by the O type stars, suggest their origin to be 
influenced by an H\,{\sc ii} region expanding into the surrounding ISM. 
All these observational evidences support the notion that the star formation triggered by feedback such as from 
H\,{\sc ii} regions may be an important mechanism for high-mass star formation \citep[e.g.,][]{deharveng10} in this particular region. 
\section{Summary and Conclusions}
\label{sec:conc}
We studied the star formation process around the bubble CN 148, 
associated with the H\,{\sc ii} region G10.3-0.1, using multi-wavelength observations. 
This work provides a careful look at various archival public data surveys (e.g. MAGPIS, ATLASGAL, Hi-GAL, MIPSGAL, GLIMPSE, UWISH2, 2MASS, and UKIDSS), 
along with previously published radio 21 cm continuum and molecular line (CO) observations of the source. 
This study used a more appropriate distance to CN 148. 
The important conclusions of this work are as follows:\\

(i) The arc-shaped distribution of molecular CO gas along with cold dust emission and PAH features trace 
a PDR around the H\,{\sc ii} region.\\

(ii) The presence of a PDR and several arc-like structures of H$_{2}$ emission, 
directly suggests the impact and influence of the H\,{\sc ii} region on its surrounding. 
The observed curvature morphology of the molecular dense material is likely caused by the expansion of the H\,{\sc ii} region. \\

(iii) High-resolution MAGPIS 20 cm radio continuum data trace three distinct 
compact components including an UCH\,{\sc ii} region G10.30-0.15, and each of them is found to be powered by early 
B type stars.\\

(iv) Hi-GAL images (70--500 $\mu$m) show a very similar morphology to the one seen in the ATLASGAL 870 $\mu$m map. 
Filament-like features are seen in the Hi-GAL images, in the direction northeast from the bubble. 
Using the ATLASGAL survey at 870 $\mu$m, 14 cold dust clumps were found with 
a total dust mass of $\sim$ 7264 M$_{\odot}$ and all of them are associated with A$_{V}$ larger than $\sim$ 16 mag. 
The detection of several high-mass as well as dense 870 $\mu$m cold dust 
continuum clumps, along with the CO emission, show fragmentation of the dense molecular material along the bubble. \\

(v) Five starless clumps are also spatially located at the extreme outer boundary of ATLASGAL 870 $\mu$m 
emission, away from the edges of the bubble.\\ 

(vi) Four 6.7 GHz methanol masers (MME1-4) were detected at the edges of the bubble, two 
of which (MME1 and MME4) are physically associated with the EGOs in the northeast (G10.34-0.14) 
and southwest (G10.29-0.13) directions. 
Previously, these EGOs were characterized as MYSOs, associated with outflow activity and lack of radio continuum emission. 
The two remaining 6.7 GHz methanol masers (MME2 and MME3) are physically linked with a massive protostar 
candidate (mass$\sim$7.7$\pm$2.4M$_{\odot}$) and the UCH\,{\sc ii} region G10.30-0.15.  These characteristics favour the 
presence of different early stages of MSF at the edges of the bubble.\\

(vii) {\it Spitzer}-IRAC and NIR photometric data revealed 371 YSOs in the region. 
41\% (i.e. 154 from a total of 371 YSOs) of these YSOs are present in 13 
clusters, with A$_{V}$ larger than $\sim$ 16 mag, and are spatially distributed in the northeast to southwest directions. 
The density distribution of all YSOs is well correlated with that of the molecular gas and cold dust emission in the PDR.\\

(viii) The surface density of Class I and Class II YSOs reveals that the clusters at the edges of 
the bubble (both northeast and southwest) are relatively younger than the clusters located further away from the bubble. \\

(ix) The distribution of molecular gas, cold dust emission, 6.7 GHz masers, EGOs, starless clumps, 
and an apparent age spread in YSO clusters provide an argument for the existence of 
star formation activity around the bubble, which is triggered by the expanding H\,{\sc ii} region.\\

Based on all the observational evidences presented in this paper, we suggest that the bubble CN 148 is a site of possible triggered star 
formation, resulting from the expansion of the H\,{\sc ii} region. 
\section*{Acknowledgments}
We thank the anonymous referee for a critical reading of the manuscript and several useful comments and 
suggestions, both for the present manuscript and from earlier submissions to MNRAS journal, which greatly improved 
the scientific contents of the paper. 
This work is based on data obtained as part of the UKIRT Infrared Deep Sky Survey and UWISH2 survey. This publication 
made use of data products from the Two Micron All Sky Survey (a joint project of the University of Massachusetts and 
the Infrared Processing and Analysis Center / California Institute of Technology, funded by NASA and NSF), archival 
data obtained with the {\it Spitzer} Space Telescope (operated by the Jet Propulsion Laboratory, California Institute 
of Technology under a contract with NASA). The ATLASGAL project is a collaboration between the Max-Planck-Gesellschaft, the 
European Southern Observatory (ESO) and the Universidad de Chile. 
LKD is supported by the grant SFRH/BPD/79741/2011, from FCT (Portugal). 
We thank Dirk Froebrich for providing the narrow-band H$_{2}$ image through 
UWISH2 survey. We also thank Henrik Beuther for providing the integrated molecular $^{13}$CO(2-1) fits file. 

\end{document}